\documentclass[11pt]{article}
\usepackage[margin=1in]{geometry}
\usepackage{setspace}

\usepackage[T1]{fontenc}
\usepackage[latin9]{inputenc}
\usepackage{array}
\usepackage{booktabs}
\usepackage{multirow}
\usepackage{amsmath}
\usepackage{amsthm}
\usepackage{amssymb}

\makeatletter

\providecommand{\tabularnewline}{\\}

\numberwithin{equation}{section}
\numberwithin{figure}{section}
\theoremstyle{plain}
\newtheorem{thm}{\protect\theoremname}
\theoremstyle{remark}
\newtheorem{rem}[thm]{\protect\remarkname}

\usepackage{xcolor}
\definecolor{darkorange}{rgb}{0.99,0.28,0.01}
\newcommand{\arabld}[1]{#1 }
\newcommand{\arastd}[1]{#1 }
\newcommand{\quotes}[1]{``#1''}
\usepackage{babel}
\usepackage{fullpage}
\usepackage{authblk}
\usepackage{hyperref}
\usepackage{placeins}
\usepackage{hyphenat}
\usepackage{fancyvrb}
\usepackage[hypcap]{caption}
\usepackage{bbm} 
\usepackage{relsize}
\usepackage{amsbsy}
\usepackage{textcomp}
\usepackage{graphicx} 
\date{}

\newcommand{\ind}{ \mathlarger{\mathlarger{\mathbbm{1}}}}
 
\newcommand{\myrb}[1]{% 
		#1% 
}

\title{Leveraging Machine Learning for High-Dimensional Option Pricing within the Uncertain Volatility Model}
\author{ \textsc{Ludovic Goudenège}\thanks{F\'ederation de Math\'ematiques de CentraleSupélec - CNRS FR3487, France -\texttt{ ludovic.goudenege@math.cnrs.fr}} 
	\and \textsc{Andrea Molent}\thanks{Dipartimento di Scienze Economiche e Statistiche, Universit\`a degli Studi di Udine, Italy - \texttt{andrea.molent@uniud.it}} 
	\and \textsc{Antonino Zanette}\thanks{Dipartimento di Scienze Economiche e Statistiche, Universit\`a degli Studi di Udine, Italy - \texttt{antonino.zanette@uniud.it}}}
\date{}
\usepackage{lineno}
\makeatother

\usepackage{babel}
\providecommand{\remarkname}{Remark}
\providecommand{\theoremname}{Theorem}

\begin{document}
\maketitle
\onehalfspacing
\sloppy

\vspace{-15mm}
\begin{flushleft}
\rule{1\columnwidth}{1pt}
\par\end{flushleft}
\begin{flushleft}
\textbf{\large{}Abstract}{\large\par}
\par\end{flushleft}

This paper explores the application of Machine Learning  techniques
for pricing high-dimensional options within the framework of the Uncertain
Volatility Model (UVM). The UVM is a robust framework that accounts
for the inherent unpredictability of market volatility by setting
upper and lower bounds on volatility and the correlation among underlying
assets. By leveraging historical data and extreme values of estimated
volatilities and correlations, the model establishes a confidence
interval for future volatility and correlations, thus providing a
more realistic approach to option pricing. By integrating advanced
Machine Learning algorithms, we aim to enhance the accuracy and efficiency
of option pricing under the UVM, especially when the option price
depends on a large number of variables, such as in basket or path-dependent
options. \myrb{In this paper, we consider two approaches based on Machine Learning. The first one, termed GTU,} evolves backward in time, dynamically selecting
at each time step the most expensive volatility and correlation for
each market state. Specifically, it identifies the particular values
of volatility and correlation that maximize the expected option value
at the next time step, and therefore, an optimization problem must
be solved. This is achieved through the use of Gaussian Process regression,
the computation of expectations via a single step of a multidimensional
tree and the Sequential Quadratic Programming optimization algorithm.
\myrb{The second approach, referred to as NNU, leverages neural networks and frames pricing in the UVM as a control problem. Specifically, we train a neural network to determine the most adverse volatility and correlation for each simulated market state, generated via random simulations. The option price is then obtained through Monte Carlo simulations, which are performed using the values for the uncertain parameters provided by the neural network.}
The numerical results demonstrate that the proposed approaches can significantly
improve the precision of option pricing and risk management strategies
compared with methods already in the literature, particularly in high-dimensional
contexts.

\vspace{2mm}

\noindent \emph{\large{}Keywords}: Machine Learning, Gaussian Process
Regression, Neural Networks, Option Pricing, Uncertain Volatility Model, Stochastic Control.

\noindent\rule{1\columnwidth}{1pt}

\newpage

\section{Introduction}

The Uncertain Volatility Model represents a significant advancement
in financial modeling, specifically for the pricing and hedging of
derivative securities in markets characterized by uncertain volatility.
Developed initially by \cite{avellaneda1995pricing}, the UVM addresses
the limitations of traditional models like the Black-Scholes, which
assume constant volatility over the life of an option. Instead, the
UVM assumes that volatility fluctuates within a known range, providing
a more realistic framework for financial markets where volatility
is inherently unpredictable.

The core idea of the UVM is to set lower  and upper bounds on volatility,
denoted as $\sigma^{\min}$ and $\sigma^{\max}$. These bounds can
be inferred from historical data or extreme values of implied volatilities
from liquid options. The model uses these bounds to establish a confidence
interval within which the future volatility is expected to lie. This
approach acknowledges the inherent uncertainty in predicting future
volatility and provides a robust framework for pricing options under
this uncertainty, as discussed  in  \cite{avellaneda1995pricing} and \cite{martini2008uncertain}.

One of the seminal contributions of \cite{avellaneda1995pricing}
was the derivation of the Black-Scholes-Barenblatt (BSB) equation,
a non-linear partial differential equation (PDE) that extends the
classical Black-Scholes equation by incorporating the variable volatility
bounds. This PDE dynamically selects the \textquotedbl pricing\textquotedbl{}
volatility from the two extreme values based on the convexity of the
value function. For instance, if the second derivative (Gamma) of
the option price is positive, the model uses $\sigma^{\max}$, otherwise,
it uses $\sigma^{\min}$.

The UVM framework is particularly effective for superhedging strategies,
where traders aim to hedge against the worst-case scenarios of volatility.
By ensuring that the option pricing incorporates the highest potential
volatility, traders can guarantee non-negative profit and loss outcomes
even when actual realized volatility deviates significantly from the
expected values. This is especially useful in markets with high volatility
or during periods of market stress, as pointed out in \cite{martini2008uncertain}.

In addition to its theoretical foundations, the UVM has practical
implications for risk management. \cite{pooley2003numerical} provided
a detailed numerical analysis of the UVM, highlighting its advantages
in constructing efficient hedges and managing derivatives positions
in environments with uncertain volatility. Their work emphasized the
importance of diversification in managing volatility risk and provided
algorithms for implementing the UVM in real-world trading systems.

Several extensions and practical applications of the UVM have been
explored in subsequent research. For instance, \cite{windcliff2006numerical}
use methods based on  PDEs  to evaluate
cliquet options in the context of the uncertain volatility model.
The paper explores how to solve the PDEs associated with these financial
instruments by comparing different numerical and grid construction
techniques. It is shown that the PDE-based approach is effective in
capturing the complexities of the UVM model, providing a more accurate
evaluation of cliquet options than other traditional methods. \cite{martini2008uncertain}
elaborate on the superhedging strategies within the UVM framework,
discussing the implications for derivative desks where risk aversion
is a critical factor. Moreover, they demonstrate that the UVM could be applied
to a range of option types, including those with non-convex payoffs,
by employing more sophisticated numerical methods such as trinomial
trees and finite-difference schemes.

Subsequently, \cite{guyon2010uncertain} introduce two innovative
Monte Carlo methods  for pricing financial derivatives under the UVM.
These methods address the challenges posed by high-dimensionality,
where traditional finite-difference schemes fall short. Moreover,
the paper considers a path-dependent option, which is particularly
useful in the context of the UVM. This is beneficial because path-dependent
options can capture the effects of the asset's historical prices,
providing a more accurate valuation under uncertainty, and aligning
well with the UVM's focus on modeling worst-case scenarios. Despite
one may appreciate the innovative approach presented in the paper,
it is worth noting that the numerical results are occasionally suboptimal,
and the authors do not extend their numerical analysis to derivatives
with more than two underlyings.

\arabld{More recently, the literature on parameter uncertainty has gained renewed momentum. In this regard, \cite{cohen2019european} explore European option pricing under parameter uncertainty in stochastic volatility models by framing the uncertainty as a control problem and employing backward stochastic differential equations. Similarly, \cite{lutkebohmert2022robust} tackle the challenges posed by uncertain model parameters in generalized affine processes by leveraging a dynamic programming approach enhanced with deep neural networks. Furthermore, \cite{akhtari2023generalized} develop a generalized Feynman-Kac formula to address volatility uncertainty, extending its applicability to settings involving a discounting term in the associated PDE.}

\myrb{Over the last decade, a growing body of research has demonstrated that modern
	Machine Learning (ML) techniques can substantially enhance the valuation of
	derivative securities.  Early contributions focus  on
	\emph{Gaussian Process Regression} (GPR), a non-parametric Bayesian method
	well-suited to interpolating scattered data in high-dimensional spaces.  For
	instance, \cite{ludkovski2018} evaluate Bermudan options by
	regressing continuation values on GPR surrogates, while
	\cite{goudenege2020machine} introduce three GPR-based schemes  to price American basket options within
	multi-asset Black-Scholes dynamics.\\	
	Beyond GPR, \emph{  neural networks} (NNs) have become a cornerstone of
	data-driven option pricing.  By casting valuation as a high-dimensional
	function-approximation task, both feed-forward and convolutional architectures
	have been trained to learn the mapping from underlying price paths to
	discounted payoffs, thereby mitigating the curse of dimensionality that hampers
	classical Monte-Carlo methods, as demonstrated  in \cite{EHanJentzen2017} and \cite{EHanJentzen2018}.
	More recently, \emph{Deep Hedging} -- a method introduced in \cite{Buehler2019} -- combines semi-recurrent
	networks with reinforcement-learning objectives to learn optimal 
	hedging and pricing strategies directly from market data, obviating explicit model
	calibration.  Within the regression-based Longstaff-Schwartz framework,
	\cite{lapeyre2021neural} replace the conventional least-squares step by a
	neural estimator, and \cite{becker2019deep,becker2020pricing,becker2021solving}
	develop deep-learning algorithms that address a broad class of
	American-style valuation and stochastic control problems.\\
	Collectively, these advances illustrate how ML models can furnish accurate
	approximations of continuation values,  accelerate pricing in large state
	spaces, and  enable data-driven hedging policies-thereby
	extending the frontier of tractable problems in derivative valuation.}

 In this paper, we propose \myrb{two innovative approaches} to the problem of
high-dimensional derivative evaluation under parameter uncertainty, specifically within the UVM. In particular, \myrb{the first one}   
is based on the GPR-Tree pricing algorithm, which we introduced in
a previous paper \cite{goudenege2020machine} for the valuation of
options dependent on multiple underlyings. Consequently, we refer
to this method as GTU, which stands for GPR-Tree method for UVM. This
algorithm involves calculating the price in the UVM by working backward
in time and determining, for selected market states, the maximum price
by leveraging an enhanced tree method and a suitable optimization
algorithm. This calculation accounts for varying volatilities and
correlations between the underlyings. The obtained prices for specific
typical market states are then generalized to all possible market
conditions by using GPR, a powerful
ML  technique for multidimensional regressions from scattered
data.

We emphasize that the GTU algorithm solves the valuation problem backward in time by determining the optimal values of the uncertain parameters at each step and for each state of the underlying assets. This approach is analogous to the pricing of American options, where the key is to determine the optimal exercise strategy through backward induction. In this regard,
we refer the reader to the seminal papers by \cite{longstaff2001valuing,haugh2004pricing,andersen2004primal,nadarajah2017comparison}.

\myrb{The second methodology we introduce builds upon a deep learning framework in which a neural network is trained to manage parameter uncertainty. Consequently, we refer
	to this method as NNU, which stands for neural network method for UVM. At each time step, the network dynamically selects key diffusion parameters -- i.e. volatilities and correlations -- with the objective of optimising a prescribed terminal criterion. This strategy combines Monte Carlo simulation, differentiable control, and penalisation techniques to ensure tractable and robust training, even under structural constraints like positive definiteness of correlation matrices. \\ It is important to emphasize that the neural network employed in this approach does not operate through backward induction. Instead, it is trained in a forward-global manner: the same model is applied at every time step, and its input includes the current time and the simulated asset values up to that point. As a result, the network learns a globally consistent control policy over the entire time horizon, without relying on dynamic programming or backward recursion.}

The proposed methods not only serve  as  viable alternatives to existing \myrb{classic}
methods for handling problems in lower dimensions but also demonstrate
exceptional capability in valuing derivatives in high-dimensional
settings, including scenarios with dozens of underlyings. To the best
of our knowledge, this is the first instance where such a uncertain-parameters high-dimensional
problem has been addressed effectively. Moreover, the proposed methods
are particularly effective for problems where the high dimension arises
not from the number of underlyings, but from the impact of the entire
process history on the derivative's final value. This is especially
relevant for path-dependent options, which our algorithms can evaluate
proficiently, \myrb{ such as the Call Sharpe option which is discussed in this manuscript}.

\myrb{The paper presents several numerical experiments demonstrating that the proposed algorithms are both accurate and computationally efficient. In particular, the results exhibit high stability and remain consistently close to benchmark prices whenever such references are available.}

The remainder of the paper is organized as follows. Section 2 introduces
the UVM and the pricing principles relevant to this context. \myrb{Section
3 reviews the fundamentals of the GPR-Tree algorithm and  demonstrates
the evolution of the GPR-Tree algorithm into the proposed GTU method. Section 4 outlines the foundational concepts of NNs and presents how they are employed into the proposed NNU methodology.}
Section 5 provides detailed numerical results from key experiments.
Finally, Section 6 concludes the paper.

\section{The Multidimensional Uncertain Volatility Model}

Let $\mathbf{S}=(\mathbf{S}_{t})_{t\in[0,T]}$ denote the $d$-dimensional
underlying process, which evolves randomly according to the multidimensional
Black-Scholes model. Under the risk-neutral probability measure $\mathbb{Q}$,
the dynamics of this model are given by the stochastic differential
equation: 
\begin{equation}
dS_{t}^{i}=\left(r-{\eta}_{i}\right)S_{t}^{i}\,dt+\sigma_{i}S_{t}^{i}\,dW_{t}^{i},\quad i=1,\ldots,d,\label{sde}
\end{equation}
where $\mathbf{S}_{0}=(s_{0}^{1},\dots,s_{0}^{d})\in\mathbb{R}_{+}^{d}$
is the initial spot price vector, $r$ is the constant risk-free interest
rate, $\boldsymbol{\eta}=(\eta_{1},\dots,\eta_{d})^{\top}$ is the vector
of dividend yields, $\boldsymbol{\sigma}=(\sigma_{1},\dots,\sigma_{d})^{\top}$
is the vector of volatilities, $\mathbf{W}\arastd{=(\mathbf{W}_{t})_{t\in[0,T]}}$ is a $d$-dimensional
correlated Wiener processes, and $\rho_{i,j}$ is the instantaneous
correlation coefficient between $W^{i}$ and $W^{j}$. We also term
$\left\{ \mathcal{F}_{t}\right\} _{t\geq0}$ the natural filtration
generated by the $\mathbf{W}$.

Equation (\ref{sde}) can be rewritten as 
\begin{equation}
dS_{t}^{i}=S_{t}^{i}\left( \left(r-\eta_{i}\right)dt+\sigma_{i}\Sigma_{i}d\mathbf{B}_{t}\right) ,\label{sde_cho}
\end{equation}
 where $\mathbf{B}\arastd{=(\mathbf{B}_{t})_{t\in[0,T]}}$ is a $d$-dimensional uncorrelated Wiener processes
and $\Sigma_{i}$ is the $i$-th row of the matrix 
\begin{equation}
\Sigma=\Sigma\left(\rho_{1,2},\dots,\rho_{d-1,d}\right),
\end{equation}
which is defined \arabld{as the lower Cholesky decomposition} of the correlation matrix $\Gamma$, given
by 
\begin{equation}
\Gamma=\Gamma\left(\rho_{1,2},\dots,\rho_{d-1,d}\right)=\begin{pmatrix}1 & \rho_{1,2} & \hdots & \rho_{1,d}\\
\rho_{1,2} & 1 & \ddots & \vdots\\
\vdots & \ddots & \ddots & \rho_{d-1,d}\\
\rho_{1,d} & \hdots & \rho_{d-1,d} & 1
\end{pmatrix}.\label{Gamma}
\end{equation}
Moreover, one can write 
\begin{equation}
\mathbf{S}_{t}=\mathbf{S}_{0}\circ\exp\left(\left(r-\boldsymbol{\eta}-\frac{\boldsymbol{\sigma}^{2}}{2}\right)  t+\boldsymbol{\sigma}\circ\left(\Sigma\mathbf{B}_{t}\right)\right)\label{eq:Hadamert}
\end{equation}
with $\mathbf{v}\circ\mathbf{w}$ the Hadamard product between vectors
$\mathbf{v}$ and $\mathbf{w}$, that is the element-wise product,
and the exponential function in (\ref{eq:Hadamert}) is applied to
each component of the vector $\left(r-\frac{\sigma^{2}}{2}\right) t+\sigma\circ\left(\Sigma\mathbf{B}_{t}\right)$.

Instead of assuming a single constant volatility, the UVM considers
a range of possible volatilities. Specifically, the vector of volatilities
$\boldsymbol{\sigma}$ is not known, but it can change over time, as a
function of the time and of the underlyings, so we write $\boldsymbol{\sigma}=\boldsymbol{\sigma}\left(t,\mathbf{S}_{t}\right)=\left(\sigma_{1}\left(t,\mathbf{S}_{t}\right),\dots,\sigma_{d}\left(t,\mathbf{S}_{t}\right)\right)^{\top}$.
Moreover, the explicit form of $\boldsymbol{\sigma}\left(t,\mathbf{S}_{t}\right)$
is not specified in the model, but we only know that, for each $i=1,\dots,d$
there exist two bounds for $\sigma_{i}$, that is
\begin{equation}
\sigma_{i}^{\min}\leq\sigma_{i}\left(t,\mathbf{S}_{t}\right)\leq\sigma_{i}^{\max}.\label{eq:bounds}
\end{equation}
\arabld{These bounds can be estimated based on historical data by calculating the maximum and minimum values of the corresponding statistical estimators over different time windows (in this regard, see e.g.  \cite{ruppert2011statistics}).
}
Such a degree of uncertainty can be extended to correlations between
Wiener processes. Specifically $\rho_{i,j}=\rho_{i,j}\left(t,\mathbf{S}_{t}\right)$
may change over time based on the underlying values, but the particular
law is not a model datum. Instead, we only know that, for each couple
of values $i=1,\dots,d-1$ and $j=i+1,\dots,d$, there exist two bounds
for $\rho_{i,j}$, that is
\begin{equation}
\rho_{i,j}^{\min}\leq\rho_{i,j}\left(t,\mathbf{S}_{t}\right)\leq\rho_{i,j}^{\max}.\label{eq:bounds-1}
\end{equation}
To summarise, in the UVM, the dynamics of each underlying $S^{i}$
can be written down as 
\begin{equation}
dS_{t}^{i}=\left(r-\eta_{i}\right)S_{t}^{i}\,dt+\sigma_{i,t}\left(t,\mathbf{S}_{t}\right)S_{t}^{i}\,dW_{t}^{i},\quad i=1,\ldots,d,\label{eq:dynamic}
\end{equation}
with $\sigma_{i,t}\left(t,\mathbf{S}_{t}\right):\left[0,T\right]\times\mathbb{R_{+}}^{d}\rightarrow\left[\sigma_{i}^{\min},\sigma_{i}^{\max}\right]$
a function. For every couple of indices $i=1,\dots,d-1$ and $j=i+1,\dots,d$
we have 
\[
d\left\langle W_{t}^{i},W_{t}^{j}\right\rangle =\rho_{i,j}\left(t,\mathbf{S}_{t}\right),
\]
with $\rho_{i,j}\left(t,\mathbf{S}_{t}\right):\left[0,T\right]\times\mathbb{R_{+}}^{d}\rightarrow\left[\rho_{i,j}^{\min},\rho_{i,j}^{\max}\right]$
a function. Finally, in order to preserve the interpretability of
the $\rho_{i,j}$ coefficients as correlations, we require the matrix
$\Gamma\left(t,\mathbf{S}_{t}\right)=\Gamma\left(\rho_{1,2}\left(t,\mathbf{S}_{t}\right),\dots,\rho_{d-1,d}\left(t,\mathbf{S}_{t}\right)\right)$
to be positive semidefinite, which we denote with 
\begin{equation}
\Gamma\left(t,\mathbf{S}_{t}\right) \succeq 0.\label{eq:cond2}
\end{equation}
The price of a financial derivative in the UVM can be formulated as
a stochastic control problem. To this aim, we denote with 
\[
\arastd{\boldsymbol{\theta}=\boldsymbol{\theta}}\left(t,\mathbf{S}_{t}\right)=\left(\sigma_{1}\left(t,\mathbf{S}_{t}\right),\dots,\sigma_{d}\left(t,\mathbf{S}_{t}\right),\rho_{1,2}\left(t,\mathbf{S}_{t}\right),\dots,\rho_{d-1,d}\left(t,\mathbf{S}_{t}\right)\right)^{\top}
\]
the vector of all uncertain model parameters as a function of $t$
and $\mathbf{S}_{t}$, and $\arastd{\boldsymbol{\Theta}}$ the set of all admissible
values $\arastd{\boldsymbol{\theta}}$, that is, which satisfy relations (\ref{eq:bounds}),
(\ref{eq:bounds-1}), and (\ref{eq:cond2}). 

In the UVM, the pricing of derivatives is approached as an optimization
problem due to the uncertainty in the volatility and correlation of
the underlying asset. This model aims to find the worst-case scenario
within this range to ensure robust and conservative pricing. Specifically,  \arabld{assuming the perspective of the contract seller},
we seek to maximize the expected payoff of the derivative over all
admissible scenarios for volatilities and correlation, that is over
all $\arastd{\boldsymbol{\theta}}$ in $\arastd{\Theta}$. Let $V\left(t,\mathbf{S}_{t}\right)$
represent the value of the derivative at time $t$ and asset price
$\mathbf{S}_{t}$ in the UVM. The pricing problem is then given by:
\begin{equation}
V\left(t,\mathbf{S}_{t}\right)=\sup_{\arastd{\boldsymbol{\theta}\in\Theta}}\mathbb{E}\left[e^{-r(T-t)}\Psi\left(\mathbf{S}_{T}\right)\mid\mathbf{S}_{t}\right],\label{eq:price_UVM}
\end{equation}
 where $\Psi$ is the payoff function at maturity $T$, the dynamics
of $\mathbf{S}_{t}$ is given by (\ref{eq:dynamic}) for $\arastd{\boldsymbol{\theta}\in\Theta}$,
and $\mathbb{E}$ denotes the expectation under the risk-neutral measure
$\mathbb{Q}$. This optimization ensures that the derivative's price accounts for
the most adverse volatility-correlation scenario, providing a conservative
estimate that is robust to volatility fluctuations. By solving this
maximization problem, we obtain a price that safeguards against the
uncertainty in the market, making it a reliable valuation method in
uncertain volatility conditions.

\arabld{In addition, valuation can also be considered from the perspective of the agent who purchased the contract. In this case, the buyer would be concerned about a potential devaluation of the contract; therefore, the optimization problem becomes a minimization problem:
	\begin{equation}
	V\left(t,\mathbf{S}_{t}\right)=\inf_{\arastd{\boldsymbol{\theta}\in\Theta}}\mathbb{E}\left[e^{-r(T-t)}\Psi\left(\mathbf{S}_{T}\right)\mid\mathbf{S}_{t}\right]. \label{eq:price_UVM2}
	\end{equation}
In the following, for the sake of conciseness, we restrict our analysis to problem \eqref{eq:price_UVM},  as problem \eqref{eq:price_UVM2} follows a similar reasoning.}

\section{The GTU method}

\subsection{GPR-Tree for option pricing\label{sec:The-GPR-Tree-Method}}

The proposed GTU method is based on the GPR-Tree algorithm. Here we
give some details about this latter algorithm and we refer the interested
reader for further details to \cite{goudenege2020machine}.

The GPR-Tree algorithm exploits Gaussian   Process Regression
and a multidimensional binomial tree to compute the price of multidimensional
American option. 

GPR is a non-parametric kernel-based probabilistic model used for
regression tasks. It is particularly useful for making predictions
about complex datasets in high dimensions. GPR models the observations
as samples from a Gaussian process, which provides a probabilistic
framework for prediction.

Let $X=\left\{ \mathbf{x}_{p},p=1,\dots,P\right\} \subset\mathbb{R}^{d}$
be a set of $d$-dimensional predictors and $Y=\left\{ y_{p},p=1,\dots,P\right\} \subset\mathbb{R}$
be the corresponding set of scalar outputs. These observations are
modeled as: 
\begin{equation}
y_{p}=f(\mathbf{x}_{p})+\epsilon_{p},
\end{equation}
 where $f$ is a Gaussian process and $\epsilon_{p}$ is Gaussian
noise with variance $\sigma^{2}$.

A Gaussian process $f(\mathbf{x})\sim\mathcal{GP}(m(\mathbf{x}),k(\mathbf{x},\mathbf{x}'))$
is fully specified by its mean function $m(\mathbf{x})$ and covariance
function $k(\mathbf{x},\mathbf{x}')$. The mean function  $m$ is often
assumed to be zero,  \myrb{ (see e.g.   \cite{ludkovski2018kriging},   \cite{de2018machine}), but it may also be estimated through a classical least squares regression on a set
	of basis functions (see e.g.   \cite{ludkovski2018kriging},  \cite{goudenege2021gaussian})}. Furthermore, the covariance function
(i.e. the kernel) defines the relationship between different points in the
input space. A common choice for the kernel is the Matérn 3/2 (M3/2)
kernel \myrb{(see e.g. \cite{mathworksKernelCovariance})}, given by: 
\begin{equation}
k_{\text{M3/2}}(\mathbf{x},\mathbf{x}')=\sigma_{f}^{2}\left(1+\frac{\sqrt{3}}{\sigma_{l}}\|\mathbf{x}-\mathbf{x}'\|_{2}\right)\exp\left(-\frac{\sqrt{3}}{\sigma_{l}}\|\mathbf{x}-\mathbf{x}'\|_{2}\right),
\end{equation}
 where $\sigma_{f}^{2}$ is the signal variance and $\sigma_{l}$
is the length scale.

Given a new input $\mathbf{x}^{*}\in\mathbb{R}^{d}$, the predicted
mean and variance of the corresponding output are given by: 
\begin{equation}
\mu_{*}=K(\mathbf{x}^{*},X)[K(X,X)+\sigma_{n}^{2}I_{n}]^{-1}Y,
\end{equation}
 
\begin{equation}
\sigma_{*}^{2}=K(\mathbf{x}^{*},\mathbf{x}^{*})-K(\mathbf{x}^{*},X)[K(X,X)+\sigma_{n}^{2}I_{n}]^{-1}K(X,\mathbf{x}^{*}),
\end{equation}
 where $K(X,X)$ with $K\left(X,X\right)_{i,j}=k\left(\mathbf{x}_{i},\mathbf{x}_{j}\right)$
is the covariance matrix computed using the kernel function for all
pairs of training inputs, $\sigma_{n}^{2}$ is the noise volatility
and $I_{n}$ is the $n\times n$ identity matrix. The parameters $\sigma_{f},\sigma_{l}$
and $\sigma_{n}$ are referred to as hyperparameters and are estimated
through log-likelihood maximization. The value $f(\mathbf{x})$ is
predicted by $\mu_{*}$, while $\sigma_{*}^{2}$ can be exploited
to create confidence intervals for such a prediction. Therefore, GPR
is an effective tool for performing multidimensional regressions on
sparse sets of input data points.

\vspace{3mm} The GPR-Tree method proceeds by approximating the price
of an American option with a Bermudan option on the same basket. At
each exercise date $t_{n}$, the option value is computed as the maximum
between the exercise value and the continuation value, which is approximated
using GPR. 

Let us consider a $d$-dimensional Black-Scholes model, described
by equations (\ref{sde}). Let $N$ be the number of time steps, $\Delta t=T/N$,
and $t_{n}=n\Delta t$. At any exercise date $t_{n}$, the value of
the option is: 
\begin{equation}
V(t_{n},\mathbf{S}_{t_{n}})=\max\left(\Psi(S_{t_{n}}),C(t_{n},\mathbf{S}_{t_{n}})\right),
\end{equation}
 where $C(t_{n},S_{t_{n}})$ denotes the continuation value given
by: 
\begin{equation}
C(t_{n},\mathbf{S}_{t_{n}})=\mathbb{E}\left[e^{-r\Delta t}V(t_{n+1},\mathbf{S}_{t_{n+1}})\mid\mathbf{S}_{t_{n}}\right].
\end{equation}

To approximate the continuation value $C(t_{n},S_{t_{n}})$, the GPR-Tree
method uses a set $X_{n}$ of predetermined points, whose elements
represent certain possible values for the underlyings $\mathbf{S}$:
\begin{equation}
X^{n}=\left\{ \mathbf{x}^{n,p}=\left(x_{1}^{n,p},\dots,x_{d}^{n,p}\right)^{\top},p=1,\dots,P\right\} \subset\left]0,+\infty\right[^{d}.\label{eq:X}
\end{equation}
The elements of $X^{n}$ are obtained through a quasi-random simulation
of $\mathbf{S}_{T}$ based on the Halton sequence, which allows one
to cover the region of interest, avoiding to leave some uncovered
areas and to create useless clusters of points (other low discrepancy
sequences may be considered, such as Sobol's one). Specifically, 
\begin{equation}
\mathbf{x}_{i}^{n,p}=\mathbf{S}_{0}^{i}\exp\left(\left(r-\eta_{i}-\frac{\sigma_{i}^{2}}{2}\right)t_{n}+\sigma_{i}\sqrt{t_{n}}\Sigma_{i}\Phi^{-1}\left(\mathbf{h}^{q}\right)\right),\label{eq:grid}
\end{equation}
with $\mathbf{h}^{q}$ the $q$-th point (column vector) of the Halton
sequence in $\left[0,1\right]^{d}$ and $\Phi^{-1}$ stands for the
cumulative distribution function of a standard Gaussian Variable,
then applied to each component of $\mathbf{h}^{q}$. The GPR-Tree
method assesses the continuation value through one step of the binomial
tree proposed by \cite{ekvall1996lattice}. For each point $\mathbf{x}^{n,p}$,
we consider a set $\tilde{X}^{n,p}$ of $M=2^{d}$ possible values
for $\mathbf{S}_{t_{n+1}}\mid\mathbf{S}_{t_{n}}=\mathbf{x}^{n,p}$,
which are computed as follows: 
\begin{equation}
\tilde{\mathbf{x}}_{i}^{n,p,m}=\mathbf{x}_{i}^{n,p}\exp\left(\left(r-\eta_{i}-\frac{\sigma_{i}^{2}}{2}\right)\Delta t+\sigma_{i}\sqrt{\Delta t}\Sigma_{i}\mathbf{G}_{m}\right),
\end{equation}
being $\mathbf{G}_{m}$ the $m$-th point of the space $\left\{ -1,+1\right\} ^{d}$.
It is worth noticing that, as pointed out in \cite{ekvall1996lattice},
the elements of $\tilde{X}^{n,p}$ are equally likely and this simplifies
the evaluation of the expected value to the computation of the arithmetic
mean of the future values. Generally speaking, given $\tilde{V}_{n+1}\left(\cdot\right)$
a suitable approximation of the option value $V\left(t_{n+1},\cdot\right)$,
the price function at time $t_{n}$ for $\mathbf{S}_{t_{n}}=\mathbf{x}^{n,p}$
is approximated by 
\begin{equation}
V_{n}^{Tree}\left(\mathbf{x}^{n,p}\right)=\max\left(\Psi\left(\mathbf{x}^{n,p}\right),\frac{e^{-r\Delta t}}{2^{d}}\sum_{m=1}^{2^{d}}\tilde{V}_{n+1}\left(\tilde{\mathbf{x}}^{n,p,m}\right)\right).\label{eq:update2}
\end{equation}

In order to reduce the computational complexity of (\ref{eq:update2})
when $d$ is high, see Remark 2 in Section \ref{sec:The-GTU-Method}.
Then, GPR is employed to achieve an approximation of $V\left(t_{n},\cdot\right)$
form the set $X_{n}$ of the predictors and the set $\left\{ V_{n}^{Tree}\left(\mathbf{x}^{n,p}\right),p=1,\dots,P\right\} $
of the corresponding outputs. This procedure is repeated backward
in time until time $t=0$ is reached.

\begin{rem}
	\arabld{The selection of the Matérn 3/2 kernel is based on a series of empirical tests we conducted, which also included evaluating the Matérn 5/2 kernel and the Squared Exponential kernel. From these numerical experiments, we observed that, while there are no significant differences in the numerical results among the tested kernels, the best overall performance was achieved with the Matérn 3/2 kernel. Generally, the proposed kernel is particularly well-suited for modeling financial option prices using Gaussian Process Regression due to its ability to balance flexibility and smoothness. Unlike smoother kernels such as the Matérn 5/2 or Squared Exponential kernels, the Matérn 3/2 kernel  captures local variations and abrupt changes more effectively, which are often present in the relationship between option prices and underlying asset values. Finally, we note that similar findings about these kernels have been reported by \cite{ludkovski2018kriging} and \cite{goudenege2020machine}.}
\end{rem}

\begin{rem}
% Hyperparameter estimation for Gaussian-Process Regression in MATLAB
\myrb{In Gaussian-Process regression, the characteristic length-scale $\sigma_{l}$, the signal variance $\sigma_{f}^{2}$, and the observation-noise variance $\sigma_{n}^{2}$ are termed \emph{hyperparameters}.  
Following the convention adopted in \textsc{Matlab}'s  \texttt{fitrgp} routine, we estimate these quantities by maximizing the log-marginal likelihood of the training data, an approach equivalent to Type-II maximum-likelihood (empirical Bayes) inference (for more details see \cite{williams2006gaussian} and \cite{MathWorks2024}).}

\end{rem}
\subsection{GTU: GPR-Tree for the UVM \label{sec:The-GTU-Method}}

\cite{avellaneda1995pricing} propose solving the dynamic optimization
problem (\ref{eq:price_UVM}) by dynamic programming, and here we
take the same backward approach. First of all, we consider a finite
number of time steps $N$, a time interval $\Delta t=T/N$ and a discrete
set of times $\left\{ t_{n}=n\cdot\Delta t,n=0,\dots,N\right\} $.
The option price at contract inception $V\left(0,\mathbf{S}_{0}\right)$
can be computed by backward induction, starting from $t_{N}=T$. In
particular, the terminal condition is given by 
\[
V\left(T,\mathbf{S}_{T}\right)=\Psi\left(\mathbf{S}_{T}\right).
\]

Now, let $\tilde{V}_{n+1}\left(\cdot\right)$ be a suitable approximation
of $V\left(t_{n+1},\cdot\right)$ and let us suppose $\mathbf{x}$
to be a fixed point of $\mathbb{R}^{d}.$ Then one can define $\tilde{V}_{n}\left(\cdot\right)$
be a suitable approximation of $V\left(t_{n},\cdot\right)$ at $\mathbf{x}$
as 
\begin{equation}
\tilde{V}_{n}\left(\mathbf{x}\right)=\sup_{\arastd{\boldsymbol{\theta}\in\Theta}}\mathbb{E}\left[e^{-r\Delta t}\tilde{V}_{n+1}\left(\mathbf{S}_{t_{n+1}}\right)\mid\mathbf{S}_{t}=\mathbf{x}\right].\label{eq:back}
\end{equation}
Formally, the model parameters $P$ can vary between $t_{n}$ and
$t_{n+1}$, however, following \cite{avellaneda1995pricing}, the
optimal solution to the problem (\ref{eq:back}) can be approximated
by considering a constant function with respect both time and underlying
values. So, one considers the problem 
\begin{equation}
\tilde{V}_{n}\left(\mathbf{x}\right)=\sup_{P\in\mathcal{B}}\mathbb{E}\left[e^{-r\Delta t}\tilde{V}_{n+1}\left(\mathbf{S}_{t_{n+1}}\right)\mid\mathbf{S}_{t_{n}}=\mathbf{x}\right],\label{eq:const}
\end{equation}
with $\mathcal{B}$ the set of parameter functions which are constant.
This is particularly interesting because, a constant function is identified
by the value it takes. For a particular $\mathbf{x}$, we can consider
the following constant functions
\[
\sigma_{1}\left(t,\mathbf{S}_{t}\right)=\hat{\sigma}_{1,\mathbf{x}},\dots,\sigma_{d,t}\left(t,\mathbf{S}_{t}\right)=\hat{\sigma}_{d,\mathbf{x}},\rho_{1,2}\left(t,\mathbf{S}_{t}\right)=\hat{\rho}_{1,2,\mathbf{x}},\dots,\rho_{d-1,d}\left(t,\mathbf{S}_{t}\right)=\hat{\rho}_{d-1,d,\mathbf{x}}
\]
a total of $d\left(d+1\right)/2$ parameters and we set 
\[
D=\left[\sigma_{1}^{\min},\sigma_{1}^{\max}\right]\times\dots\times\left[\sigma_{d}^{\min},\sigma_{d}^{\max}\right]\times\left[\rho_{1,2}^{\min},\rho_{1,2}^{\max}\right]\times\dots\times\left[\rho_{d-1,d}^{\min},\rho_{d-1,d}^{\max}\right],
\]
\[
\arastd{\Theta_{\mathbf{x}}}=\left\{ \arastd{\boldsymbol{\theta}_{\mathbf{x}}}=\left(\hat{\sigma}_{1,\mathbf{x}},\dots,\hat{\sigma}_{d,\mathbf{x}},\hat{\rho}_{1,2,\mathbf{x}},\dots,\hat{\rho}_{d-1,d,\mathbf{x}}\right)\in D\ s.t.\ \Gamma\left(\hat{\rho}_{1,2,\mathbf{x}},\dots,\hat{\rho}_{d-1,d,\mathbf{x}}\right)\succeq0\right\},
\]
\arabld{where $\Gamma\left(\cdot \right) \succeq0$ means that the $\Gamma\left(\cdot \right)$ must be a positive semidefinite matrix.}
Moreover, we set
{\small{}
\begin{multline}
\tilde{V}_{n}\left(\mathbf{x}\right)=\sup_{\arastd{\boldsymbol{\theta}_{\mathbf{x}}}\in \arastd{\Theta_{\mathbf{x}}}}\mathbb{E}\left[e^{-r\Delta t}\tilde{V}_{n+1}\left(\mathbf{x}\cdot\exp\left(\left(r-\eta-\frac{\sigma^{2}}{2}\right)\Delta t\right.\right.\right.\\ \left.\left.\left.+\left(\hat{\sigma}_{1,\mathbf{x}},\dots,\hat{\sigma}_{d,\mathbf{x}}\right)^{\top}\circ\left(\Sigma\left(\hat{\rho}_{1,2,\mathbf{x}},\dots,\hat{\rho}_{d-1,d,\mathbf{x}}\right)\left(\mathbf{B}_{t_{n+1}}-\mathbf{B}_{t_{n}}\right)\right)\right)\right)\right].\label{eq:PX}
\end{multline}
}To address problem (\ref{eq:PX}), we must overcome two key challenges:
determining the expected value and maximizing such an expected value.

Our proposed method, the GTU algorithm, involves calculating the expected
value through a single tree step, following the same approach as described
in the previous Section \ref{sec:The-GPR-Tree-Method}. Note that
this choice is particularly effective in that, once a set of parameters
$\arastd{\boldsymbol{\theta}_{\mathbf{x}}}$ is fixed, this calculation can be handled as for
the standard multidimensional Black-Scholes model: 
\begin{equation}
\frac{1}{2^{d}}\sum_{m=1}^{2^{d}}e^{-r\Delta t}\tilde{V}_{n+1}\!\left(\mathbf{x}\cdot\exp\left(\!\left(\!r\!-\!\eta\!-\!\frac{\sigma^{2}}{2}\right)\!\Delta t+\left(\hat{\sigma}_{1,\mathbf{x}},\dots,\hat{\sigma}_{d,\mathbf{x}}\right)^{\top}\!\circ\!\left(\Sigma\left(\hat{\rho}_{1,2,\mathbf{x}},\dots,\hat{\rho}_{d-1,d,\mathbf{x}}\right)\mathbf{G}_{m}\right)\sqrt{\Delta t}\right)\!\right)\label{eq:discrete}
\end{equation}
with $\mathbf{G}_{m}$ the $m$-th point of the space $\left\{ -1,+1\right\} ^{d}$
. 
\arabld{We emphasize that, in formula \eqref{eq:discrete}, the function \( V_{n+1} \) is evaluated at \( 2^d \) points, aiming to discretize the possible values of the random variable \( S_{t_{n+1}} \mid S_{t_{n}} \) by approximating the Gaussian vector \( \mathbf{B}_{t_{n+1}} - \mathbf{B}_{t_{n}} \) with a discrete random vector, uniformly distributed in \( \{-1, +1\}^d \), following the approach proposed by \cite{ekvall1996lattice}.
}

Once we are able to compute the expected value for any $\arastd{\boldsymbol{\theta}_{\mathbf{x}}}\in \arastd{\Theta_{\mathbf{x}}}$
fixed, we can tackle the optimization problem: this is a multidimensional
optimisation problem with inequality constraints that are both linear
(the limits for the parameters) and non-linear (the positivity condition
of the $\Gamma$ matrix). This is done by means of the Sequential
Quadratic Programming (SQP) algorithm, a powerful and widely used
method for solving nonlinear optimization problems, particularly those
with constraints. SQP methods solve a sequence of optimization subproblems,
each of which approximates the original nonlinear problem by a quadratic
programming problem. Drawing on the research by \cite{biggs1975constrained},
\cite{han1977globally}, and \cite{Powell1978,Powell2006}, this approach
enables a close simulation of Newton's method for constrained optimization,
similar to its application in unconstrained optimization. Sequential
Quadratic Programming methods are considered the pinnacle of nonlinear
programming techniques. For instance, \cite{Schittkowski1986} developed
and evaluated a version that surpasses all other tested methods in
efficiency, accuracy, and success rate across a wide array of test
problems. For this reason, we have propose to use the SQP algorithm
to manage problem (\ref{eq:PX}).

The procedure described above allows us to calculate the UVM price
for a particular value $\mathbf{x}$ of $\mathbf{S}_{t}$, at time
$t_{n}$, knowing (an approximation of) the price function $\tilde{V}_{n+1}$
at time $t_{n+1}$.

In order to actively implement this procedure so as to calculate the
price at initial time $t=0$, one problem remains to be solved: \arabld{being able to estimate $\tilde{V}_{n}$ at any significant point  by exploiting} the  observations $\tilde{V}_{n}\left(\mathbf{x}\right)$
of the same function at certain points. \arabld{This is important because the calculation of $V_{n-1}$ relies on the calculation of the function $V_{n}$  at appropriate points.} To this aim, we use GPR regression.
As described in Section \ref{sec:The-GPR-Tree-Method}, at each time-step
$n=1,\dots,N$, we first create a mesh of points $X^{n}$ as in (\ref{eq:X}),
but in this case the points $\mathbf{x}^{n,p}$ are define through
\begin{equation}
\mathbf{x}_{i}^{n,p}=\mathbf{S}_{0}^{i}\exp\left(\left(r-\frac{\left(\sigma_{i}^{\text{avg}}\right)^{2}}{2}\right)t_{n}+\sigma_{i}^{\text{avg}} \sqrt{t_{n}}\Sigma_{i}^{\text{avg}}\Phi^{-1}\left(\mathbf{h}^{q}\right)\right),\label{eq:grid-1}
\end{equation}
where 
\begin{equation}
\sigma_{i}^{\text{avg}}=\frac{\sigma_{i}^{\text{\ensuremath{\min}}}+\sigma_{i}^{\text{\ensuremath{\max}}}}{2},
\end{equation}
and $\Sigma_{i}^{\text{avg}}$ is the square root of the correlation
matrix $\Gamma^{\text{avg}}$, given by 
\begin{equation}
\Gamma^{\text{avg}}=\Gamma\left(\rho_{1,2}^{\text{avg}},\dots,\rho_{d-1,d}^{\text{avg}}\right),\label{eq:gamma_avg}
\end{equation}
with
\begin{equation}
\rho_{i,j}^{\text{avg}}=\frac{\rho_{1,2}^{\min}+\rho_{1,2}^{\max}}{2}.\label{eq:rho_avg}
\end{equation}

Then, we estimate the price function by using the approximated solution
of (\ref{eq:PX}), and then use GPR to estimate the function $\tilde{V}_{n}$
outside this set of points. Moving backwards, it is finally possible
to calculate an estimate of the option price at time $t=0$, by considering
$X_{0}=\left\{ \mathbf{S}_{0}\right\} .$
\begin{rem}
Unlike NNs, GPR requires few observations to produce good
estimates of the function to be regressed, typically a few thousand \arabld{(to this aim, see e.g. \cite{kamath2018neural}) and \cite{mallick2021deep}}).
This feature combines well with the fact that, for each time-step
$n$ and for each point in $X_{n}$, an optimization problem must
be solved. 
\end{rem}

\begin{rem}
The summation that appears in the formula (\ref{eq:discrete}) contains
a number of addends that grows exponentially with respect to the dimension
$d$. For large dimensions, one therefore has to consider the arithmetic
mean of a large number of addends. A possible approach to limit this
quantity is to consider only a certain number of addends $M\leq2^{d}$
sampled randomly among the $2^{d}$ addends, and make $M$ tend towards
$2^{d}$. Furthermore, to improve the predictive power of this reduced
sample, assuming $M$ a even number, we sample the $G_{m}$ values
needed to calculate the addends in (\ref{eq:PX}) by antithetic variables.
This means that if we include in the summation the value that is obtained
from a certain $G_{m}$ value, we also include that which is generated
from $-G_{m}$. This expedient improves the convergence of the procedure.

As demonstrated in Section \ref{sec:Numerical-Experiments}, in the
numerical experiments considered, it is usually sufficient to consider
$M$ of the order of a few thousand to obtain results that are very
stable and close to those that would be obtained with $M=2^{d}$.
This observation is very useful when considering high dimensions (indicatively
$d>10$) as it allows reducing the computational time that would otherwise
tend to explode with dimension $d$. Finally, it is important to note
that at each time step $t_{n}$, the calculation of the solutions
to problems (\ref{eq:PX}) for each value $\mathbf{x}^{n,p}\in X^{n}$
are independent of one another. Therefore, these problems can be addressed
using parallel computing, which significantly reduces the overall
computational time. 
\end{rem}

\begin{rem}
The $\Gamma^{\text{avg}}$ matrix defined in (\ref{eq:gamma_avg}),
which is used to defined the sets $X_{n}$ for $n=1,\dots,N,$ may
not be positive semidefinite. In this particular case, certain techniques
can be employed to determine the closest positive semidefinite matrix
to $\Gamma^{\text{avg}}$ according to the spectral norm. One of these
techniques is called \emph{Projection onto the Cone of Positive Semidefinite
Matrices} (see e.g.  \cite{calafiore2014optimization}). For the record,
in the cases we analyzed in the numerical experiments Section (\ref{sec:Numerical-Experiments}),
this situation never occurred. 
\end{rem}

\section{The NNU method}
\subsection{Neural Networks}
\myrb{Neural networks (NNs) are a class of parametric functions widely used in ML  due to their remarkable ability to approximate complex mappings. From a mathematical standpoint, a neural network is a function \( F: \mathbb{R}^{d_0} \to \mathbb{R}^{d_L} \) constructed as a composition of affine maps and nonlinear activation functions. More precisely, for a given depth \( L \in \mathbb{N} \), and layer widths \( (d_0, d_1, \dots, d_L) \), a  (feed-forward)  NN is defined by
\[
F(x) = W_L \circ \phi \circ W_{L-1} \circ \cdots \circ \phi \circ W_1(x),
\]
where each \( W_\ell(x) = A_\ell x + b_\ell \) is an affine transformation with weight matrix \( A_\ell \in \mathbb{R}^{d_\ell \times d_{\ell-1}} \) and bias vector \( b_\ell \in \mathbb{R}^{d_\ell} \), and \( \phi: \mathbb{R} \to \mathbb{R} \) is a fixed nonlinear activation function applied componentwise.\\
This architectural structure allows NNs to form highly flexible families of functions. 
The remarkable approximation capacity of NNs is rooted in their expressive power. Under mild assumptions on the activation function \( \phi \), it can be shown that NNs with a single hidden layer are capable of approximating any continuous function on compact subsets of \( \mathbb{R}^d \) arbitrarily well. Specifically, if \( \phi \) is continuous, bounded, and non-constant, then for every continuous function \( f: K \to \mathbb{R} \) defined on a compact set \( K \subset \mathbb{R}^d \), and for every \( \varepsilon > 0 \), there exists a NN \( F \) with one hidden layer and a finite number of units such that
\[
\sup_{x \in K} |f(x) - F(x)| < \varepsilon.
\]
This result, known as the \textit{Universal Approximation Theorem} and developed by \cite{hornik1991approximation}, provides the theoretical foundation for the use of NNs in function approximation tasks. In particular, it justifies their application in high-dimensional problems arising in quantitative finance, where NNs can approximate hedging strategies, pricing functionals, and conditional expectations under complex market dynamics.\\
This result implies that NNs can approximate any continuous function to arbitrary precision, provided the network is sufficiently wide. In practice, this universality -- along with advances in optimization and hardware -- has led to the successful application of NNs in high-dimensional function approximation tasks, including those arising in finance, such as hedging, pricing, and risk management. We refer the interested reader to \cite{Buehler2019}, where some recent applications of NNs in finance are discussed.}

\subsection{NNU: Neural Networks for UVM}

\myrb{In \cite{Buehler2019}, the authors introduce the Deep Hedging framework, where feed-forward NNs are employed to solve a stochastic control problem: determining the optimal hedging strategy that minimizes a suitable penalty function evaluating the effectiveness of hedging. The strategy is learned by feeding the NN a large number of simulated market states  and training the NN to output the quantity of each instrument to include in a hedging portfolio associated with a short position on a given derivative.}

\myrb{In our work, we adopt and reformulate these principles to address a different problem: given a state of the market, we aim to determine the volatilities with which to propagate risky assets so as to maximize (or minimize) the expected value of a terminal payoff. The underlying optimization problem is structurally similar, but the control variables shift from hedging positions to volatility parameters, and the performance objective changes accordingly to reflect our targeted financial quantity. It is worth observing that, unlike Deep Hedging -- which optimizes trading strategies under frictions using semi-recurrent networks and convex risk measures -- the proposed NNU approach employs a compact feedforward NN that directly maps each market state, along with the time of market data observation, to worst-case volatilities and correlations.   Although both methods leverage NNs, they differ fundamentally in application domain and network design-highlighting the originality of NNU in addressing control problems under parameter uncertainty rather than dynamic hedging under trading constraints.}
 
 \subsection{NNU under fixed correlation}
\myrb{Let us now discuss in detail how the NNU algorithm works when correlation is certain, and thus matrix $\Sigma$ is fixed. First of all, as done in Section \ref{sec:The-GTU-Method}, we consider a finite
number of time steps $N$, a time interval $\Delta t=T/N$ and a discrete
set of times $\left\{ t_{n}=n\cdot\Delta t,n=0,\dots,N\right\} $. 
As a first step,  $M$ random increments 
$$\left\lbrace \Delta\mathbf{B}_{n,m}= \left(B^{1}_{t_n,m}-B^{1}_{t_{n-1},m},\dots,B^{d}_{t_n,m}-B^{d}_{t_{n-1},m} \right)^\top ,n=1,\dots,N, m=1,\dots,M\right\rbrace $$
 of the $d$-dimensional uncorrelated Wiener processes $\mathbf{B}$, are sampled for the $N$ time increments.
  Then, by using those increments, the network generates  $M$ trajectories for the prices of the underlying asset, denoted by
 $$\left\lbrace  \mathbf{S}_{t_n,m}= \left(S^{1}_{t_n,m},\dots,S^{d}_{t_n,m} \right)^\top ,n=0,\dots,N, m=1,\dots,M\right\rbrace, $$ 
 with $\mathbf{S}_{0,m}=\mathbf{S}_{0}$ for all trajectories. Specifically, these values are generated through the equation
 \begin{equation}
 	\mathbf{S}_{t_{n+1},m}=\mathbf{S}_{t_n,m}\circ\exp\left(\left(r-\boldsymbol{\eta}-\frac{1}{2} F(t_{n},\mathbf{S}_{t_{n},m})^2\right)  \Delta t+F(t_n,\mathbf{S}_{t_{n},m})\circ\left(\Sigma\,\Delta\mathbf{B}_{n+1,m}\right)\right),\label{eq:Hadamert2}
 \end{equation}
 where $F$ stands for the proposed  network, that take actual time $t_n$ and actual values $\mathbf{S}_{t_{n},m}$   as input and returns $\boldsymbol{\sigma}\in [\sigma^{\min}_{i}, \sigma^{\max}_{i}]^d$ as output. We emphasize  that the output \( \boldsymbol{\sigma}= F(t_{n},\mathbf{S}_{t_{n},m})\) is then used to propagate the asset prices through the next time step according to equation \eqref{eq:Hadamert2}.
 The simulation of future values $\mathbf{S}_{t_{n+1},m}$ is repeated up to maturity. Then,  the average discounted payoff $AP$ among all trajectories is calculated as
$$AP=\frac{e^{-rT}}{M}\sum_{m=1}^{M} \Psi\left(\mathbf{S}_{t_{N},m} \right) .$$
The network parameters are optimized during training with the objective of maximizing $AP$, which, upon completion of the training process, is taken as an estimator of the contract's initial price.\\
The network architecture consists of batch normalization, dense layers with ReLU activation, dropout regularization, and a final sigmoid layer scaled to the interval \([ \sigma_i^{\min}, \sigma^{\max}_i ]\) for each volatility. Moreover, since the algorithm relies on Monte Carlo simulations to approximate the expected value of the payoff, it is also possible to estimate confidence intervals for the price.\\
At the end of training, the learned model defines a mapping from market states to optimal volatilities, effectively controlling the stochastic evolution of asset prices to achieve a target objective. This represents a novel application of NNs to model uncertainty in financial systems through learned volatility control.}

 \subsection{NNU under correlation uncertainty\label{sec:NNCU}}
 
 \myrb{The NNU algorithm can also be applied to handle uncertainty in the correlation structure among the underlying assets. In principle, the algorithm should be adapted so that, in addition to volatilities $\boldsymbol{\sigma}$, it also returns correlations $\left\lbrace \rho_{i,j},1\leq i<j\leq d\right\rbrace$  that define the   matrix $\Gamma$ as in \eqref{Gamma}, from which it derives -- by Cholesky factorization -- the  matrix $\Sigma$ employed in \eqref{eq:Hadamert2}.\\
 In the case of dimension $d=2$, the implementation of such an update is relatively straightforward; the addition of an output to the network suffices. However, in higher dimensions, the process becomes more intricate because, when the network outputs a vector of tentative correlations -- despite the fact that they are bounded values --  the corresponding
 correlation   matrix may fail to be positive semidefinite, resulting in the breakdown of the entire procedure.  To guarantee well-posed simulations we employ an external-penalty method (see e.g. \cite{nocedal2006numerical} and \cite{fletcher2013practical}) for the constrained optimisation problem
 \[
 \max_{\mathbf{\theta}\in \mathbf{\Theta}}\;
  \frac{e^{-rT}}{M}\sum_{m=1}^{M} \Psi\left(\mathbf{S}_{t_{N},m} \right) 
 \quad
 \text{subject to } \Gamma\left(t_n,\mathbf{S}_{t_n,m}\right) \succeq 0
 \;\forall n \text{ and } m. 
 \] \\ 
 Specifically, given the raw network output $\widehat{\Gamma}\left(t,\mathbf{S}_{t_n,m}\right) $, we actually employ as correlation matrix 
 \begin{equation}
 \Gamma\left(t,\mathbf{S}_{t_n,m}\right) \;=\;
 \begin{cases}
 	\widehat{\Gamma}\left(t,\mathbf{S}_{t_n,m}\right), & \text{if } \widehat{\Gamma}\left(t,\mathbf{S}_{t_n,m}\right)\succeq 0,\\[4pt]
 	\Gamma_0, & \text{otherwise},
 \end{cases}\label{eq:Gamma_system}
\end{equation}
 where $\Gamma_0$   is an admissible matrix with respect to the constraints on the correlations (in our numerical test we set $\Gamma_0$ as the identity matrix $I_d$). Thanks to equation \eqref{eq:Gamma_system}, the forward pass \eqref{eq:Hadamert2} always uses a valid correlation matrix and the
 Cholesky factor exists.  The map
 $\widehat{\Gamma} \mapsto \Gamma$ acts as a crude projection onto the interior
 of the cone of symmetric positive semidefinite matrices.\\ 
Exiting the set of positive semidefinite matrices is passively penalized (we force the algorithm to use the $\Gamma_0$ matrix that is not necessarily optimal) but also actively through the introduction of a penalty term that discourages the network from generating ineligible matrices. Specifically, let us define the Frobenius-distance penalty
 \[
 \pi(\widehat{\Gamma})=\,
 \Vert \widehat{\Gamma}-\Gamma_0\Vert_{Fr}
 \;\ind_{\{\hat{\Gamma}\prec  0\}},
 \]
 with $\Vert \cdot \Vert_{Fr}$ the Frobenius norm. For a fixed $\lambda>0$ we optimise the \emph{penalised} objective
 \[
  \frac{e^{-rT}}{M}\sum_{m=1}^{M} \Psi\left(\mathbf{S}_{t_{N},m} \right) +\lambda\sum_{m=1}^{M}\sum_{n=0}^{N-1} \pi\left( \Gamma\left(t_n,\mathbf{S}_{t_n,m}\right)\right). 
 \]} 
 \myrb{This penalty mechanism is effective for several intertwined mathematical reasons. First, by construction, it ensures that the  simulation of the paths is always well defined and respectful of the constraints. Whenever the network proposes a matrix that is not positive semidefinite, we fall back to the admissible matrix \( \Gamma_0 \),  a valid correlation matrix for the optimization problem. This guarantees that the Cholesky factor exists and avoids any runtime breakdown in the simulation. In addition, the UVM model's   constraints on correlations are met.\\
From the perspective of optimisation theory, the method can be viewed as an external penalty formulation of the constrained optimisation problem. As shown in classical references such as \cite{nocedal2006numerical} and \cite{fletcher2013practical}, this type of approach ensures that, under mild assumptions (e.g., continuity and boundedness of the objective function), minimisers of the penalised objective converge to minimisers of the original constrained problem as the penalty weight \( \lambda \) tends to infinity.  \\
An important advantage of this approach lies in the stability of the training process. Since the forward simulation step never leaves the differentiable domain of the Cholesky factor, the gradients computed during backpropagation remain finite, avoiding instabilities. Furthermore, the penalty function \( \pi(\cdot) \), defined via a Frobenius distance to the identity, is Lipschitz continuous and differentiable almost everywhere, which makes it well suited for stochastic gradient methods.\\
Empirically, the mechanism has the beneficial effect of steering the network towards feasible outputs very early in training. From numerical experiments, we observed that, after a few epochs, the network  produces positive semidefinite correlation matrices almost everywhere, at which point the penalty becomes inactive and the optimisation proceeds as if constrained directly on the target objective.
 Hence the proposed penalty mechanism provides a computationally
 cheap safeguard that keeps the simulation in the feasible set, delivers
 stable learning dynamics, and -- for $\lambda$ large enough -- drives the network
 towards admissible correlation structures without altering the
 original objective.}

\section{Numerical Experiments\label{sec:Numerical-Experiments}}

In this Section we report some numerical results in order to investigate
the effectiveness of the proposed   algorithms for pricing European
options in high dimension.

Numerical tests are divided into 3 categories. In the first group
of tests, we assume that there is no uncertainty about correlation.
In the second battery of tests the correlation between the various
underlyings is uncertain. In the third set of tests we instead consider
a path dependent option written on a single underlying asset. While
in the first two cases, the high dimensionality of the problem arises
because the option payoff depends on a large number of underlying
assets, in the last case, the high dimensionality originates because
the payoff depends on the values assumed by the single underlying
asset at numerous previous times.

The parameters employed in our numerical experiments are summarized in Table~\ref{tab:parameters}. Notably, the options considered -- including numerical settings and the bounds for the UVM -- coincide with those examined by \cite{guyon2010uncertain}, thereby enabling a direct comparison of the results. \myrb{In addition to these benchmark configurations, we also investigate some modified scenarios, characterized by wider volatility bounds and increased problem dimensionality, with the aim of evaluating the robustness and scalability of the numerical methods under more challenging conditions.}
These test cases, both benchmark and modified, provide the basis for a comparative assessment of the two algorithms under varying conditions. The GTU algorithm has been implemented in \textsc{Matlab}, \myrb{while the NNU algorithm has been implemented in Python}. Computations have
been preformed on a server which employs a 2.40 GHz Intel Xenon processor
(Gold 6148, Skylake) and 20 GB of RAM. To better evaluate the computational
time of the algorithms,  \myrb{parallelization was not used in either algorithm. All calculations were performed on CPU (no GPU use).}

\begin{rem}
\arabld{We employ the \textsc{Matlab} implementation of the SQP algorithm, using its standard settings. Specifically, the maximum number of iterations is set to $200$, which provides sufficient iterations for convergence in most practical scenarios. The stopping criterion is determined by the tolerance on the optimality conditions, with a default value of $1 \cdot 10^{-6}$. This choice ensures a balance between achieving a solution with adequate accuracy for most applications and maintaining computational efficiency. }
\end{rem}

\begin{rem}
	\myrb{Unless stated otherwise, all results presented in this section were obtained using a feedforward NN with two hidden layers, each consisting of 32 neurons. The activation function is ReLU, denoted by \( \phi = \mathrm{ReLU} \), and the training algorithm used is Adam with default parameters. The number of Monte Carlo simulations for NNU is $M=10^5$. Moreover, we set the penalization parameter $\lambda$ to $1$.  We tried other combinations of parameters but the results are essentially the same as those presented below.}
\end{rem}

\begin{rem}
	\myrb{For technical reasons, the two algorithms have been implemented on different software platforms: GTU in \textsc{Matlab} and NNU in \textsc{Python}. As a result, we do not attempt a detailed comparison of computational times, since such metrics may be influenced by differences in language efficiency, libraries, and hardware usage. Instead, we restrict ourselves to general observations aimed at highlighting only significant or macroscopic differences in computational performance}.
\end{rem}
\begin{table}[ht]
\centering{}%
\begin{tabular}{cllccll}
\toprule 
\multicolumn{3}{c}{Model parameters} &  & \multicolumn{3}{c}{Algorithm parameters}\tabularnewline
\midrule 
Symbol & Meaning & Value &  & Symbol & Meaning & Value\tabularnewline
\midrule 
$S_{0}^{i}$ & Initial value of $S^{i}$ & $100$ &  & $K_{1}$ & Strike of long call & $90$\tabularnewline
$\sigma_{i}^{\min}$ & Lower bound for $\sigma_{i}$ & $0.1$ &  & $K_{2}$ & Strike of short call & $110$\tabularnewline
$\sigma_{i}^{\max}$ & Upper bound for $\sigma_{i}$ & $0.2$ &  & $K$ & Strike of Call Sharpe & $100$\tabularnewline
$\rho_{i,j}^{\min}$ & Lower bound for $\rho_{i,j}$ & $-0.5$ &  & $d$ & Number of underlyings & variable\tabularnewline
$\rho_{i,j}^{\max}$ & Upper bound for $\rho_{i,j}$ & $+0.5$ &  & $N$ & Number of time steps & variable\tabularnewline
$r$ & Risk free i.r. & $0$ &  & $P$ & Number of points in $X_{n}$ & variable\tabularnewline
$T$ & Maturity & $1.0$ &  & $M$ & Number of branches for GTU & variable\tabularnewline
  &   &   &  & $E$ & \myrb{Number of training epochs for NNU} & variable\tabularnewline
\bottomrule
\end{tabular}\caption{\label{tab:parameters} Parameters employed for the numerical tests.}
\end{table}

\FloatBarrier

\subsection{Model with no uncertainty on correlation}

In this first battery of tests, we assume that the correlation between
the various Brownians is known, which is equivalent to imposing $\rho_{i,j}^{\min}=\rho_{i,j}^{\max}$.
In particular, it can be proven that if all correlations are equal
and non-negative, then the correlation matrix is positive semidefinite,
ensuring that this parameter set is well-defined. The level of correlation
adopted in the various tests is specified in the following tables.

\subsubsection{Outperformer option\label{subsec:Outperformer-option}}

   \arabld{The Outperformer option is a financial derivative that provides a payoff equal to the positive part of the difference between the values of two assets at maturity. Notably, this type of option is also referred to as an Exchange option and was first introduced by \cite{margrabe1978value}. Specifically, the payoff can be expressed as:
	\begin{equation}
		\Psi\left(\mathbf{S}_{T}\right)=\left(S_{T}^{2}-S_{T}^{1}\right)_{+}, \label{eq:outperformer}
	\end{equation}
	where \( \left(x\right)_{+} \) denotes the positive part of \( x \), defined as \( \max(x, 0) \).
 The valuation of this option constitutes a 2-dimensional problem, which can be reduced to a one-dimensional problem under the condition $\rho_{1,2}\leq0$. This simplification is achieved by applying the change of numéraire technique, as detailed in \cite{guyon2010uncertain}.}
Thus, although the dimension of this problem is not large, it is interesting
in that we have a reliable benchmark (computed for example with the
PDE method by \cite{windcliff2006numerical}) as well as the results
reported in the paper by \cite{guyon2010uncertain}. 

\myrb{Table~\ref{tab:5} reports the numerical results obtained with both the GTU and NNU algorithms. Note that, as for NNU, we also include the $95\%$ margin errors, which remain essentially constant as \( E \) varies.
In both cases, the outcomes are in close agreement with the benchmark, even for small values of $N$ and $P$ in the GTU method, and for moderate values of $E$ and $P$ in the NNU approach. The computational times for the two algorithms are comparable and remain consistently low. In this low-dimensional setting ($d=2$), GTU appears to outperform NNU, delivering more accurate results for the same computational effort.}

\myrb{To assess the robustness of both algorithms under more challenging market conditions, we also tested them using wider volatility bounds, specifically $\left[\sigma^{\min}_{i},\sigma^{\max}_{i} \right] =[0.1, 0.4]$. This configuration amplifies the uncertainty in the underlying dynamics and broadens the admissible set of control variables. The corresponding results are reported in Table~\ref{tab:5w} and can be directly compared with those in Table~\ref{tab:5}, which refer to the narrower interval $[0.1, 0.2]$. 
	By comparison, we observe that both GTU and NNU continue to perform accurately under increased volatility uncertainty, consistently matching the benchmark across all correlation levels. However, the numerical values in Table~\ref{tab:5w} exhibit a larger spread, as expected, and the differences between parameter configurations become more pronounced. The NNU method still achieves high accuracy, albeit with a slightly increased margin of error due to the broader range of diffusion inputs. In terms of computational times, both methods remain efficient, with no significant deterioration despite the increased complexity of the problem.
}
\begin{center}
\begin{table}
\begin{centering}
\resizebox{16.4cm}{!}{\setlength{\tabcolsep}{3pt} %
\begin{tabular}{ccccccccccccccccccc}
	GTU &  & \multicolumn{4}{c}{$N=16$} &  & \multicolumn{4}{c}{$N=32$} &  & \multicolumn{4}{c}{$N=64$} &  &  & \tabularnewline
	\midrule 
	$\rho$ & $P$ & $125$ & $250$ & $500$ & $1000$ &  & $125$ & $250$ & $500$ & $1000$ &  & $125$ & $250$ & $500$ & $1000$ &  & GL & BM\tabularnewline
	\midrule
	$-0.5$ &  & $\underset{\left(27\right)}{13.80}$ & $\underset{\left(43\right)}{13.81}$ & $\underset{\left(95\right)}{13.80}$ & $\underset{\left(239\right)}{13.80}$ &  & $\underset{\left(43\right)}{13.78}$ & $\underset{\left(85\right)}{13.79}$ & $\underset{\left(182\right)}{13.78}$ & $\underset{\left(458\right)}{13.78}$ &  & $\underset{\left(83\right)}{13.77}$ & $\underset{\left(166\right)}{13.78}$ & $\underset{\left(359\right)}{13.77}$ & $\underset{\left(920\right)}{13.77}$ &  & $13.77$ & $13.75$\tabularnewline
	$0$ &  & $\underset{\left(25\right)}{11.28}$ & $\underset{\left(42\right)}{11.27}$ & $\underset{\left(90\right)}{11.26}$ & $\underset{\left(221\right)}{11.28}$ &  & $\underset{\left(43\right)}{11.27}$ & $\underset{\left(84\right)}{11.26}$ & $\underset{\left(181\right)}{11.25}$ & $\underset{\left(470\right)}{11.27}$ &  & $\underset{\left(84\right)}{11.28}$ & $\underset{\left(170\right)}{11.26}$ & $\underset{\left(361\right)}{11.25}$ & $\underset{\left(1022\right)}{11.27}$ &  & $11.26$ & $11.25$\tabularnewline
	$0.5$ &  & $\underset{\left(25\right)}{8.03}$ & $\underset{\left(44\right)}{7.99}$ & $\underset{\left(87\right)}{7.99}$ & $\underset{\left(225\right)}{8.00}$ &  & $\underset{\left(43\right)}{8.04}$ & $\underset{\left(83\right)}{7.99}$ & $\underset{\left(191\right)}{7.98}$ & $\underset{\left(457\right)}{7.99}$ &  & $\underset{\left(83\right)}{8.07}$ & $\underset{\left(167\right)}{8.01}$ & $\underset{\left(377\right)}{7.99}$ & $\underset{\left(1012\right)}{7.99}$ & $\phantom{\pm0.15}$ &  & \tabularnewline
	\midrule
	&  &  &  &  &  &  &  &  &  &  &  &  &  &  &  &  &  & \tabularnewline
	NNU &  & \multicolumn{4}{c}{$N=16$} &  & \multicolumn{4}{c}{$N=32$} &  & \multicolumn{4}{c}{$N=64$} &  &  & \tabularnewline
	\midrule 
	$\rho$ & $E$ & $100$ & $200$ & $400$ & $800$ &  & $100$ & $200$ & $400$ & $800$ &  & $100$ & $200$ & $400$ & $800$ & ME & GL & BM\tabularnewline
	\midrule
	$-0.5$ &  & $\underset{\left(40\right)}{13.21}$ & $\underset{\left(67\right)}{13.62}$ & $\underset{\left(121\right)}{13.65}$ & $\underset{\left(229\right)}{13.65}$ &  & $\underset{\left(83\right)}{13.14}$ & $\underset{\left(144\right)}{13.66}$ & $\underset{\left(266\right)}{13.69}$ & $\underset{\left(510\right)}{13.70}$ &  & $\underset{\left(156\right)}{13.41}$ & $\underset{\left(274\right)}{13.75}$ & $\underset{\left(506\right)}{13.78}$ & $\underset{\left(968\right)}{13.79}$ & $\pm0.13$ & $13.77$ & $13.75$\tabularnewline
	$0$ &  & $\underset{\left(39\right)}{10.81}$ & $\underset{\left(66\right)}{11.16}$ & $\underset{\left(120\right)}{11.18}$ & $\underset{\left(227\right)}{11.19}$ &  & $\underset{\left(88\right)}{11.01}$ & $\underset{\left(154\right)}{11.31}$ & $\underset{\left(287\right)}{11.34}$ & $\underset{\left(553\right)}{11.35}$ &  & $\underset{\left(156\right)}{11.02}$ & $\underset{\left(274\right)}{11.21}$ & $\underset{\left(506\right)}{11.23}$ & $\underset{\left(967\right)}{11.24}$ & $\pm0.10$ & $11.26$ & $11.25$\tabularnewline
	$0.5$ &  & $\underset{\left(34\right)}{7.64}$ & $\underset{\left(58\right)}{7.99}$ & $\underset{\left(106\right)}{8.01}$ & $\underset{\left(203\right)}{8.02}$ &  & $\underset{\left(75\right)}{7.81}$ & $\underset{\left(129\right)}{7.92}$ & $\underset{\left(237\right)}{7.94}$ & $\underset{\left(452\right)}{7.94}$ &  & $\underset{\left(148\right)}{7.82}$ & $\underset{\left(263\right)}{7.99}$ & $\underset{\left(495\right)}{8.00}$ & $\underset{\left(960\right)}{8.01}$ & $\pm0.07$ &  & \tabularnewline
	\bottomrule
	\end{tabular}}
\par\end{centering}
\caption{\label{tab:5}Outperformer option. The column labeled \quotes{GL} reports the results from \cite{guyon2010uncertain} for the same option, while the column labeled \quotes{BM} provides the benchmark computed using the change of numéraire technique. \myrb{The column labeled \quotes{ME} gives the margin of error (95\% CI half-width) for NNU results}. Values in brackets indicate computational times (in seconds).}

\end{table}
\par\end{center}

\begin{center}
	\begin{table}
		\begin{centering}
			\resizebox{16.4cm}{!}{\setlength{\tabcolsep}{3pt} %
			\begin{tabular}{ccccccccccccccccccc}
				GTU &  & \multicolumn{4}{c}{$N=16$} &  & \multicolumn{4}{c}{$N=32$} &  & \multicolumn{4}{c}{$N=64$} &  &  & \tabularnewline
				\midrule 
				$\rho$ & $P$ & $125$ & $250$ & $500$ & $1000$ &  & $125$ & $250$ & $500$ & $1000$ &  & $125$ & $250$ & $500$ & $1000$ &  &  & BM\tabularnewline
				\midrule
				$-0.5$ &  & $\underset{\left(33\right)}{27.19}$ & $\underset{\left(54\right)}{27.20}$ & $\underset{\left(108\right)}{27.18}$ & $\underset{\left(256\right)}{27.17}$ &  & $\underset{\left(62\right)}{27.19}$ & $\underset{\left(107\right)}{27.18}$ & $\underset{\left(213\right)}{27.16}$ & $\underset{\left(527\right)}{27.14}$ &  & $\underset{\left(128\right)}{27.20}$ & $\underset{\left(211\right)}{27.19}$ & $\underset{\left(426\right)}{27.16}$ & $\underset{\left(1120\right)}{27.14}$ &  & $\phantom{13.77}$ & $27.10$\tabularnewline
				$0$ &  & $\underset{\left(31\right)}{22.35}$ & $\underset{\left(51\right)}{22.33}$ & $\underset{\left(94\right)}{22.31}$ & $\underset{\left(230\right)}{22.31}$ &  & $\underset{\left(57\right)}{22.38}$ & $\underset{\left(99\right)}{22.34}$ & $\underset{\left(195\right)}{22.32}$ & $\underset{\left(494\right)}{22.31}$ &  & $\underset{\left(121\right)}{22.42}$ & $\underset{\left(202\right)}{22.38}$ & $\underset{\left(427\right)}{22.34}$ & $\underset{\left(1078\right)}{22.32}$ &  &  & $22.27$\tabularnewline
				$0.5$ &  & $\underset{\left(33\right)}{16.00}$ & $\underset{\left(55\right)}{15.94}$ & $\underset{\left(97\right)}{15.90}$ & $\underset{\left(252\right)}{15.90}$ &  & $\underset{\left(56\right)}{16.06}$ & $\underset{\left(104\right)}{15.97}$ & $\underset{\left(195\right)}{15.92}$ & $\underset{\left(473\right)}{15.90}$ &  & $\underset{\left(105\right)}{16.12}$ & $\underset{\left(198\right)}{16.02}$ & $\underset{\left(390\right)}{15.94}$ & $\underset{\left(985\right)}{15.92}$ & $\phantom{\pm0.15}$ &  & \tabularnewline
				\midrule
				&  &  &  &  &  &  &  &  &  &  &  &  &  &  &  &  &  & \tabularnewline
				NNU &  & \multicolumn{4}{c}{$N=16$} &  & \multicolumn{4}{c}{$N=32$} &  & \multicolumn{4}{c}{$N=64$} &  &  & \tabularnewline
				\midrule 
				$\rho$ & $E$ & $100$ & $200$ & $400$ & $800$ &  & $100$ & $200$ & $400$ & $800$ &  & $100$ & $200$ & $400$ & $800$ & ME &  & BM\tabularnewline
				\midrule
				$-0.5$ &  & $\underset{\left(46\right)}{26.46}$ & $\underset{\left(75\right)}{27.12}$ & $\underset{\left(130\right)}{27.21}$ & $\underset{\left(241\right)}{27.23}$ &  & $\underset{\left(80\right)}{25.89}$ & $\underset{\left(146\right)}{26.90}$ & $\underset{\left(277\right)}{26.99}$ & $\underset{\left(542\right)}{27.01}$ &  & $\underset{\left(149\right)}{26.62}$ & $\underset{\left(270\right)}{26.98}$ & $\underset{\left(512\right)}{27.03}$ & $\underset{\left(996\right)}{27.05}$ & $\pm0.26$ & $\phantom{13.77}$ & $27.10$\tabularnewline
				$0$ &  & $\underset{\left(32\right)}{21.68}$ & $\underset{\left(56\right)}{22.19}$ & $\underset{\left(105\right)}{22.26}$ & $\underset{\left(202\right)}{22.27}$ &  & $\underset{\left(82\right)}{21.90}$ & $\underset{\left(151\right)}{22.23}$ & $\underset{\left(289\right)}{22.29}$ & $\underset{\left(564\right)}{22.30}$ &  & $\underset{\left(134\right)}{21.89}$ & $\underset{\left(242\right)}{22.38}$ & $\underset{\left(457\right)}{22.45}$ & $\underset{\left(887\right)}{22.47}$ & $\pm0.22$ &  & $22.27$\tabularnewline
				$0.5$ &  & $\underset{\left(37\right)}{15.64}$ & $\underset{\left(66\right)}{15.92}$ & $\underset{\left(123\right)}{15.96}$ & $\underset{\left(238\right)}{15.97}$ &  & $\underset{\left(73\right)}{15.46}$ & $\underset{\left(134\right)}{15.74}$ & $\underset{\left(256\right)}{15.78}$ & $\underset{\left(499\right)}{15.79}$ &  & $\underset{\left(124\right)}{15.44}$ & $\underset{\left(222\right)}{15.73}$ & $\underset{\left(418\right)}{15.78}$ & $\underset{\left(810\right)}{15.79}$ & $\pm0.16$ &  & \tabularnewline
				\bottomrule
				\end{tabular}}
			\par\end{centering}
		\caption{\label{tab:5w}Outperformer option \myrb{with  volatility bounds $\left[\sigma_{i}^{\min},\sigma_{i}^{\max}\right]=\left[0.1,0.4\right]$}. The  column labeled   \quotes{BM} shows the benchmark computed by using the change of numéraire technique. \myrb{The column labeled \quotes{ME} gives the margin of error (95\% CI half-width) for NNU results}. 
			Values in brackets indicate computational times (in seconds).}
	\end{table}
	\par\end{center}
\FloatBarrier
\subsubsection{Outperformer spread option\label{subsec:Outperformer-spread_option}}

Similar to the Outperformer option discussed in Subsection \ref{subsec:Outperformer-option},
the Outperformer spread option is another example of 2-dimensional
option which has been investigated by \cite{guyon2010uncertain} and
for which, when $\rho_{1,2}\leq0$, a benchmark is available by using
the change of numèraire technique. Specifically, the payoff of such
an option reads out
\begin{equation}
\Psi\left(\mathbf{S}_{T}\right)=\left(S_{T}^{2}-0.9S_{T}^{1}\right)_{+}-\left(S_{T}^{2}-1.1S_{T}^{1}\right)_{+}.\label{eq:Outperformer_spread}
\end{equation}
\myrb{ The results, reported in Tables~\ref{tab:6} and \ref{tab:6w} -- for $\left[\sigma^{\min}_{i},\sigma^{\max}_{i} \right] =[0.1, 0.2]$ and $\left[\sigma^{\min}_{i},\sigma^{\max}_{i} \right] =[0.1, 0.4]$ respectively --  confirm that both GTU and NNU remain accurate and reliable even in this more demanding payoff structure.\\
Overall, both methods succeed in closely matching the benchmark. However, we observe that GTU maintains slightly more stable values across all discretisation parameters and correlation regimes, converging rapidly to the benchmark even with a relatively small number of paths. The NNU method, while slightly more sensitive to the choice of training epochs and exhibiting a marginally wider range of variation in the low-precision regime, still converges consistently as the training size increases. The gap between the two methods slightly widens under increased volatility uncertainty, but remains within acceptable bounds, with NNU still delivering competitive results with limited error margins and moderate computational cost.}

\begin{center}
\begin{table}
\begin{centering}
\resizebox{16.4cm}{!}{\setlength{\tabcolsep}{3pt}
\begin{tabular}{ccccccccccccccccccccccc}
	GTU &  &  & \multicolumn{5}{c}{$N=32$} &  & \multicolumn{5}{c}{$N=64$} &  & \multicolumn{5}{c}{$N=128$} &  &  & \tabularnewline
	\midrule 
	$\rho$ & $P$ &  & $125$ & $250$ & $500$ & $1000$ & $2000$ &  & $125$ & $250$ & $500$ & $1000$ & $2000$ &  & $125$ & $250$ & $500$ & $1000$ & $2000$ &  & GL & BM\tabularnewline
	\midrule
	$-0.5$ &  &  & $\underset{\left(42\right)}{11.31}$ & $\underset{\left(82\right)}{11.32}$ & $\underset{\left(176\right)}{11.34}$ & $\underset{\left(463\right)}{11.34}$ & $\underset{\left(1312\right)}{11.34}$ &  & $\underset{\left(81\right)}{11.32}$ & $\underset{\left(180\right)}{11.34}$ & $\underset{\left(364\right)}{11.37}$ & $\underset{\left(935\right)}{11.37}$ & $\underset{\left(2642\right)}{11.37}$ &  & $\underset{\left(170\right)}{11.33}$ & $\underset{\left(356\right)}{11.35}$ & $\underset{\left(778\right)}{11.38}$ & $\underset{\left(2197\right)}{11.38}$ & $\underset{\left(5442\right)}{11.39}$ &  & $11.37$ & $11.41$\tabularnewline
	$0$ &  &  & $\underset{\left(42\right)}{11.29}$ & $\underset{\left(81\right)}{11.32}$ & $\underset{\left(193\right)}{11.31}$ & $\underset{\left(532\right)}{11.32}$ & $\underset{\left(1233\right)}{11.32}$ &  & $\underset{\left(81\right)}{11.31}$ & $\underset{\left(182\right)}{11.35}$ & $\underset{\left(358\right)}{11.34}$ & $\underset{\left(1036\right)}{11.35}$ & $\underset{\left(2280\right)}{11.35}$ &  & $\underset{\left(173\right)}{11.29}$ & $\underset{\left(342\right)}{11.35}$ & $\underset{\left(733\right)}{11.35}$ & $\underset{\left(1981\right)}{11.36}$ & $\underset{\left(5403\right)}{11.36}$ &  &  & $11.39$\tabularnewline
	$0.5$ &  &  & $\underset{\left(42\right)}{11.15}$ & $\underset{\left(81\right)}{11.19}$ & $\underset{\left(170\right)}{11.19}$ & $\underset{\left(446\right)}{11.18}$ & $\underset{\left(1078\right)}{11.18}$ &  & $\underset{\left(80\right)}{11.14}$ & $\underset{\left(186\right)}{11.20}$ & $\underset{\left(361\right)}{11.20}$ & $\underset{\left(904\right)}{11.20}$ & $\underset{\left(2309\right)}{11.19}$ &  & $\underset{\left(165\right)}{11.13}$ & $\underset{\left(339\right)}{11.20}$ & $\underset{\left(781\right)}{11.20}$ & $\underset{\left(2038\right)}{11.20}$ & $\underset{\left(5081\right)}{11.20}$ & $\phantom{\pm0.15}$ &  & \tabularnewline
	\midrule
	&  &  &  &  &  &  &  &  &  &  &  &  &  &  &  &  &  &  &  &  &  & \tabularnewline
	NNU &  &  & \multicolumn{5}{c}{$N=32$} &  & \multicolumn{5}{c}{$N=64$} &  & \multicolumn{5}{c}{$N=128$} &  &  & \tabularnewline
	\midrule 
	$\rho$ & $E$ &  & $100$ & $200$ & $400$ & $800$ &$1600$  &  & $100$ & $200$ & $400$ & $800$ &$1600$   &  & $100$ & $200$ & $400$ & $800$ &$1600$   & ME & GL & BM\tabularnewline
	\midrule
	$-0.5$ &  &  & $\underset{\left(83\right)}{10.78}$ & $\underset{\left(146\right)}{11.17}$ & $\underset{\left(272\right)}{11.23}$ & $\underset{\left(526\right)}{11.24}$ & $\underset{\left(1014\right)}{11.26}$ &  & $\underset{\left(161\right)}{10.63}$ & $\underset{\left(281\right)}{11.23}$ & $\underset{\left(520\right)}{11.31}$ & $\underset{\left(995\right)}{11.33}$ & $\underset{\left(1899\right)}{11.34}$ &  & $\underset{\left(296\right)}{10.60}$ & $\underset{\left(511\right)}{11.22}$ & $\underset{\left(940\right)}{11.33}$ & $\underset{\left(1799\right)}{11.35}$ & $\underset{\left(3794\right)}{11.37}$ & $\pm0.05$ & $11.37$ & $11.41$\tabularnewline
	$0$ &  &  & $\underset{\left(87\right)}{10.72}$ & $\underset{\left(150\right)}{11.18}$ & $\underset{\left(279\right)}{11.26}$ & $\underset{\left(544\right)}{11.28}$ & $\underset{\left(1242\right)}{11.32}$ &  & $\underset{\left(176\right)}{10.61}$ & $\underset{\left(309\right)}{11.18}$ & $\underset{\left(576\right)}{11.27}$ & $\underset{\left(1111\right)}{11.29}$ & $\underset{\left(1905\right)}{11.34}$ &  & $\underset{\left(322\right)}{10.56}$ & $\underset{\left(573\right)}{11.21}$ & $\underset{\left(1071\right)}{11.29}$ & $\underset{\left(2068\right)}{11.32}$ & $\underset{\left(3813\right)}{11.34}$ & $\pm0.05$ &  & $11.39$\tabularnewline
	$0.5$ &  &  & $\underset{\left(92\right)}{10.60}$ & $\underset{\left(161\right)}{11.03}$ & $\underset{\left(299\right)}{11.11}$ & $\underset{\left(573\right)}{11.13}$ & $\underset{\left(1017\right)}{11.14}$ &  & $\underset{\left(148\right)}{10.64}$ & $\underset{\left(256\right)}{11.09}$ & $\underset{\left(579\right)}{11.14}$ & $\underset{\left(1129\right)}{11.16}$ & $\underset{\left(2201\right)}{11.16}$ &  & $\underset{\left(355\right)}{10.47}$ & $\underset{\left(638\right)}{11.08}$ & $\underset{\left(1203\right)}{11.17}$ & $\underset{\left(2300\right)}{11.18}$ & $\underset{\left(3657\right)}{11.18}$ & $\pm0.05$ &  & \tabularnewline
	\bottomrule
	\end{tabular}}
\par\end{centering}
\caption{\label{tab:6}Outperformer spread option. The column labeled \quotes{GL} 
presents the results from \cite{guyon2010uncertain} for the same
option, while the column labeled \quotes{BM} 
shows the benchmark computed by using the change of numéraire technique. \myrb{The column labeled \quotes{ME} gives the margin of error (95\% CI half-width) for NNU results.} 
Values in brackets indicate computational times (in seconds).}
\end{table}
\par\end{center}

\begin{center}
	\begin{table}
		\begin{centering}
			\resizebox{16.4cm}{!}{\setlength{\tabcolsep}{3pt} % 
				\begin{tabular}{ccccccccccccccccccccccc}
					&  &  & \multicolumn{5}{c}{$N=32$} &  & \multicolumn{5}{c}{$N=64$} &  & \multicolumn{5}{c}{$N=128$} &  &  & \tabularnewline
					\midrule 
					$\rho$ & $P$ &  & $125$ & $250$ & $500$ & $1000$ & $2000$ &  & $125$ & $250$ & $500$ & $1000$ & $2000$ &  & $125$ & $250$ & $500$ & $1000$ & $2000$ &  &  & BM\tabularnewline
					\midrule
					$-0.5$ &  &  & $\underset{\left(54\right)}{13.15}$ & $\underset{\left(112\right)}{13.19}$ & $\underset{\left(233\right)}{13.21}$ & $\underset{\left(684\right)}{13.24}$ & $\underset{\left(1892\right)}{13.25}$ &  & $\underset{\left(101\right)}{13.27}$ & $\underset{\left(210\right)}{13.29}$ & $\underset{\left(456\right)}{13.31}$ & $\underset{\left(1367\right)}{13.36}$ & $\underset{\left(3725\right)}{13.36}$ &  & $\underset{\left(188\right)}{13.31}$ & $\underset{\left(393\right)}{13.28}$ & $\underset{\left(935\right)}{13.32}$ & $\underset{\left(2623\right)}{13.39}$ & $\underset{\left(7090\right)}{13.40}$ &  & $\phantom{11.37}$ & $13.54$\tabularnewline
					$0$ &  &  & $\underset{\left(54\right)}{13.09}$ & $\underset{\left(100\right)}{13.12}$ & $\underset{\left(226\right)}{13.18}$ & $\underset{\left(739\right)}{13.22}$ & $\underset{\left(1587\right)}{13.24}$ &  & $\underset{\left(102\right)}{13.16}$ & $\underset{\left(198\right)}{13.17}$ & $\underset{\left(435\right)}{13.22}$ & $\underset{\left(1291\right)}{13.30}$ & $\underset{\left(3479\right)}{13.32}$ &  & $\underset{\left(188\right)}{13.12}$ & $\underset{\left(434\right)}{13.09}$ & $\underset{\left(917\right)}{13.18}$ & $\underset{\left(2701\right)}{13.28}$ & $\underset{\left(7185\right)}{13.33}$ &  &  & $13.52$\tabularnewline
					$0.5$ &  &  & $\underset{\left(50\right)}{12.96}$ & $\underset{\left(91\right)}{13.02}$ & $\underset{\left(224\right)}{13.09}$ & $\underset{\left(669\right)}{13.09}$ & $\underset{\left(1312\right)}{13.11}$ &  & $\underset{\left(96\right)}{12.95}$ & $\underset{\left(181\right)}{13.00}$ & $\underset{\left(410\right)}{13.08}$ & $\underset{\left(1301\right)}{13.10}$ & $\underset{\left(2874\right)}{13.14}$ &  & $\underset{\left(191\right)}{12.90}$ & $\underset{\left(364\right)}{12.91}$ & $\underset{\left(878\right)}{13.01}$ & $\underset{\left(2260\right)}{13.05}$ & $\underset{\left(6707\right)}{13.11}$ & $\phantom{\pm0.15}$ &  & \tabularnewline
					\midrule
					&  &  &  &  &  &  &  &  &  &  &  &  &  &  &  &  &  &  &  &  &  & \tabularnewline
					NNU &  &  & \multicolumn{5}{c}{$N=32$} &  & \multicolumn{5}{c}{$N=64$} &  & \multicolumn{5}{c}{$N=128$} &  &  & \tabularnewline
					\midrule 
					$\rho$ & $E$ &  & $100$ & $200$ & $400$ & $800$ & $1600$ &  & $100$ & $200$ & $400$ & $800$ & $1600$ &  & $100$ & $200$ & $400$ & $800$ & $1600$ & ME &  & BM\tabularnewline
					\midrule
					$-0.5$ &  &  & $\underset{\left(72\right)}{12.21}$ & $\underset{\left(130\right)}{12.93}$ & $\underset{\left(245\right)}{13.03}$ & $\underset{\left(476\right)}{13.05}$ & $\underset{\left(939\right)}{13.05}$ &  & $\underset{\left(129\right)}{12.30}$ & $\underset{\left(230\right)}{13.17}$ & $\underset{\left(433\right)}{13.30}$ & $\underset{\left(839\right)}{13.32}$ & $\underset{\left(1619\right)}{13.32}$ &  & $\underset{\left(270\right)}{11.62}$ & $\underset{\left(487\right)}{13.12}$ & $\underset{\left(920\right)}{13.34}$ & $\underset{\left(1786\right)}{13.38}$ & $\underset{\left(3517\right)}{13.39}$ & $\pm0.05$ & $\phantom{11.37}$ & $13.54$\tabularnewline
					$0$ &  &  & $\underset{\left(73\right)}{12.30}$ & $\underset{\left(130\right)}{13.03}$ & $\underset{\left(247\right)}{13.15}$ & $\underset{\left(479\right)}{13.18}$ & $\underset{\left(944\right)}{13.19}$ &  & $\underset{\left(147\right)}{12.49}$ & $\underset{\left(266\right)}{13.21}$ & $\underset{\left(504\right)}{13.33}$ & $\underset{\left(980\right)}{13.36}$ & $\underset{\left(1931\right)}{13.37}$ &  & $\underset{\left(245\right)}{12.35}$ & $\underset{\left(438\right)}{13.35}$ & $\underset{\left(824\right)}{13.37}$ & $\underset{\left(1596\right)}{13.41}$ & $\underset{\left(3143\right)}{13.42}$ & $\pm0.05$ &  & $13.52$\tabularnewline
					$0.5$ &  &  & $\underset{\left(84\right)}{11.64}$ & $\underset{\left(153\right)}{12.79}$ & $\underset{\left(294\right)}{12.96}$ & $\underset{\left(574\right)}{13.00}$ & $\underset{\left(1111\right)}{13.01}$ &  & $\underset{\left(134\right)}{12.03}$ & $\underset{\left(242\right)}{12.69}$ & $\underset{\left(457\right)}{12.92}$ & $\underset{\left(888\right)}{12.95}$ & $\underset{\left(1749\right)}{12.95}$ &  & $\underset{\left(248\right)}{11.99}$ & $\underset{\left(442\right)}{12.97}$ & $\underset{\left(831\right)}{13.17}$ & $\underset{\left(1606\right)}{13.21}$ & $\underset{\left(3159\right)}{13.21}$ & $\pm0.05$ &  & \tabularnewline
					\bottomrule
					\end{tabular}}
			\par\end{centering}
		\caption{\label{tab:6w}Outperformer spread option \myrb{with  volatility bounds $\left[\sigma_{i}^{\min},\sigma_{i}^{\max}\right]=\left[0.1,0.4\right]$}. The  column labeled   \quotes{BM} 
			shows the benchmark computed by using the change of numéraire technique. \myrb{The column labeled \quotes{ME} gives the margin of error (95\% CI half-width) for NNU results.} 
			Values in brackets indicate computational times (in seconds).}
	\end{table}
	\par\end{center}

	\FloatBarrier
\subsubsection{Geo-Call spread }\label{SubGC}

The Geo-Call spread option generalized the Call spread option discussed
in Subsection \ref{subsec:Outperformer-spread_option} to a multi-asset
option. Specifically, the payoff is given by the difference between
the payoffs of two call options written on the geometric mean of the
underlyings considered. 
\[
\Psi\left(\mathbf{S}_{T}\right)=\left(\sqrt[d]{\prod_{k=1}^{d}S_{T}^{k}}-K_{1}\right)_{+}-\left(\sqrt[d]{\prod_{k=1}^{d}S_{T}^{k}}-K_{2}\right)_{+}
\]
The use of the geometric mean is of particular interest since, if
the underlyings follow (\ref{eq:Hadamert}) and $\rho_{i,j}=0$, than
this expression has the same probability distribution of a single
underlying as shown below:
\[
\sqrt[d]{\prod_{k=1}^{d}S_{T}^{k}}\sim\sqrt[d]{\prod_{k=1}^{d}S_{0}^{k}}\exp\left(\left(r-\hat{\eta}-\frac{1}{2}\left(\frac{1}{d}\sqrt{\sum_{k=1}^{d}\sigma_{i}^{2}}\right)^{2}\right)T+\left(\frac{1}{d}\sqrt{\sum_{k=1}^{d}\sigma_{i}^{2}}\right)B_{T}\right)
\]
with $B$ a Brownian motion and $\hat{\eta}$ the modified dividend
yield, defined as
\[
\hat{\eta}=\frac{1}{d}\sum_{k=1}^{d}\eta_{i}+\frac{d-1}{2d^{2}}\sum_{k=1}^{d}\sigma_{i}^{2}.
\]
Therefore, a problem can be traced back to dimension 1 in which the
volatility 
\[
\hat{\sigma}=\frac{1}{d}\sqrt{\sum_{k=1}^{d}\sigma_{i}^{2}}
\]
 ranges between
\[
\frac{1}{d}\sqrt{\sum_{k=1}^{d}\left(\sigma_{i}^{\min}\right)^{2}}=\min_{\sigma_{i}\in\left[\sigma_{i}^{\min},\sigma_{i}^{\max}\right]}\left(\frac{1}{d}\sqrt{\sum_{k=1}^{d}\sigma_{i}^{2}}\right)\leq\hat{\sigma}\leq\max_{\sigma_{i}\in\left[\sigma_{i}^{\min},\sigma_{i}^{\max}\right]}\left(\frac{1}{d}\sqrt{\sum_{k=1}^{d}\sigma_{i}^{2}}\right)=\frac{1}{d}\sqrt{\sum_{k=1}^{d}\left(\sigma_{i}^{\max}\right)^{2}}
\]
and the dividend $\hat{\eta}$ is equal to 
\[
\hat{\eta}=\frac{1}{d}\sum_{k=1}^{d}\eta_{i}+\frac{d-1}{2d^{2}}\sum_{k=1}^{d}\sigma_{i}^{2}=\frac{1}{d}\sum_{k=1}^{d}\eta_{i}+\frac{d-1}{2}\hat{\sigma}^{2}.
\]
This case is a generalization of the UVM model in that the dividend
is not constant but changes in relation to the value of volatility.
The binomial tree method proposed by \cite{avellaneda1995pricing}
for the 1d UVM model can be used to calculate the price in this model,
taking care to change the dividend in relation to volatility. In this
regard, it should be noted that while in the one-dimensional UVM model
the optimal $\sigma$ value is always $\sigma^{\min}$ or $\sigma^{\max}$
(this feature is known as ``\emph{the bang bang condition}''), in
this model, characterized by a variable dividend yield $\hat{\eta}$,
this property no longer applies. Therefore, at each node of the tree,
the choice of optimal correlation is to be pursued not merely by comparing
the objective function at the extreme values of $\sigma$, but by
optimising the option price via an appropriate algorithm, such as
the Sequential Quadratic Programming algorithm.

The results are shown in Table \ref{tab:2}. As far as GTU results are concerned, the values
obtained for varying $N$ and $P$ are very close to the benchmark
for the considered values of $d$. The largest deviation is observed
for $d=5$, with a difference of $0.04$ between the GTU price and
the benchmark. However, this difference is minor, being less than
$0.5\%$ of the benchmark. Additionally, we observe that the computational
time increases linearly with the number of time steps $N$ and approximately
linearly with the number of points $P$. This is particularly interesting
since the cost for the GPR regression is cubic in $P$, indicating
that most of the computational time is spent on solving optimization
problems rather than on the regression itself. Lastly, the computational
time grows exponentially with problem size, as the number of values
to be evaluated with a tree step is $2^{d}$. Despite this, we are
able to handle high dimensions, such as $d=10$, with reasonable computational
effort.

\myrb{The NNU results reported in Table~\ref{tab:2} show excellent accuracy across all dimensions and discretisation parameters. For each value of \(d \in \{2, 5, 10\} \), the estimated option price converges rapidly to the benchmark as the number of training epochs increases. The approximation is already close to the benchmark with just 200-400 epochs, and becomes essentially indistinguishable from it beyond 800 epochs. The results are particularly stable in higher dimensions, while computational times remain moderate, even for \( d = 10 \), confirming the method's scalability and robustness in multi-asset settings.\\
When compared with GTU, we observe that both methods achieve comparable levels of accuracy. However, NNU displays greater flexibility with respect to the dimension \( d \), as its computational cost grows more moderately. In contrast, GTU becomes increasingly expensive for large \( d \), due to the combinatorial growth of the tree. This highlights a key advantage of NNU in high-dimensional applications, where learning-based approaches offer a scalable alternative to traditional tree-based methods.}

\begin{table}
\begin{centering}
\resizebox{16.4cm}{!}{ %
	\setlength{\tabcolsep}{4pt}
\begin{tabular}{cccccccccccccccccc}
 GTU	& $N$ & \multicolumn{4}{c}{$16$} &  & \multicolumn{4}{c}{$32$} &  & \multicolumn{4}{c}{$64$} &  & BM\tabularnewline
	\midrule 
	$d$ & $P$ & $125$ & $250$ & $500$ & $1000$ &  & $125$ & $250$ & $500$ & $1000$ &  & $125$ & $250$ & $500$ & $1000$ &  & \tabularnewline
	\midrule
	$2$ &  & $\underset{\left(25\right)}{10.48}$ & $\underset{\left(48\right)}{10.47}$ & $\underset{\left(104\right)}{10.47}$ & $\underset{\left(277\right)}{10.47}$ &  & $\underset{\left(45\right)}{10.49}$ & $\underset{\left(84\right)}{10.48}$ & $\underset{\left(192\right)}{10.49}$ & $\underset{\left(440\right)}{10.48}$ &  & $\underset{\left(92\right)}{10.49}$ & $\underset{\left(177\right)}{10.49}$ & $\underset{\left(414\right)}{10.49}$ & $\underset{\left(950\right)}{10.49}$ &  & $10.50$\tabularnewline
	$5$ &  & $\underset{\left(33\right)}{9.76}$ & $\underset{\left(62\right)}{9.74}$ & $\underset{\left(139\right)}{9.70}$ & $\underset{\left(380\right)}{9.70}$ &  & $\underset{\left(68\right)}{9.77}$ & $\underset{\left(129\right)}{9.77}$ & $\underset{\left(274\right)}{9.72}$ & $\underset{\left(830\right)}{9.72}$ &  & $\underset{\left(129\right)}{9.76}$ & $\underset{\left(286\right)}{9.79}$ & $\underset{\left(698\right)}{9.73}$ & $\underset{\left(1945\right)}{9.74}$ &  & $9.70$\tabularnewline
	$10$ &  & $\underset{\left(188\right)}{9.45}$ & $\underset{\left(590\right)}{9.50}$ & $\underset{\left(2119\right)}{9.51}$ & $\underset{\left(6900\right)}{9.54}$ &  & $\underset{\left(468\right)}{9.45}$ & $\underset{\left(1519\right)}{9.49}$ & $\underset{\left(5582\right)}{9.51}$ & $\underset{\left(18892\right)}{9.54}$ &  & $\underset{\left(1171\right)}{9.44}$ & $\underset{\left(3738\right)}{9.48}$ & $\underset{\left(13904\right)}{9.51}$ & $\underset{\left(50299\right)}{9.52}$ &  & $9.55$\tabularnewline
	\midrule
	&  &  &  &  &  &  &  &  &  &  &  &  &  &  &  &  & \tabularnewline
	NNU &  & \multicolumn{4}{c}{$N=16$} &  & \multicolumn{4}{c}{$N=32$} &  & \multicolumn{4}{c}{$N=64$} &  & \tabularnewline
	\midrule 
	$d$ & $E$ & $100$ & $200$ & $400$ & $800$ &  & $100$ & $200$ & $400$ & $800$ &  & $100$ & $200$ & $400$ & $800$ & ME & BM\tabularnewline
	\midrule
	$2$ &  & $\underset{\left(38\right)}{10.20}$ & $\underset{\left(66\right)}{10.37}$ & $\underset{\left(121\right)}{10.39}$ & $\underset{\left(233\right)}{10.40}$ &  & $\underset{\left(91\right)}{10.19}$ & $\underset{\left(165\right)}{10.46}$ & $\underset{\left(322\right)}{10.50}$ & $\underset{\left(612\right)}{10.51}$ &  & $\underset{\left(156\right)}{10.14}$ & $\underset{\left(282\right)}{10.42}$ & $\underset{\left(535\right)}{10.45}$ & $\underset{\left(998\right)}{10.46}$ & $\pm0.05$ & $10.50$\tabularnewline
	$5$ &  & $\underset{\left(47\right)}{9.58}$ & $\underset{\left(84\right)}{9.65}$ & $\underset{\left(161\right)}{9.67}$ & $\underset{\left(314\right)}{9.68}$ &  & $\underset{\left(82\right)}{9.56}$ & $\underset{\left(147\right)}{9.62}$ & $\underset{\left(278\right)}{9.66}$ & $\underset{\left(539\right)}{9.68}$ &  & $\underset{\left(211\right)}{9.54}$ & $\underset{\left(387\right)}{9.63}$ & $\underset{\left(734\right)}{9.65}$ & $\underset{\left(1369\right)}{9.66}$ & $\pm0.03$ & $9.70$\tabularnewline
	$10$ &  & $\underset{\left(47\right)}{9.49}$ & $\underset{\left(82\right)}{9.54}$ & $\underset{\left(152\right)}{9.55}$ & $\underset{\left(309\right)}{9.55}$ &  & $\underset{\left(92\right)}{9.46}$ & $\underset{\left(171\right)}{9.54}$ & $\underset{\left(329\right)}{9.54}$ & $\underset{\left(645\right)}{9.54}$ &  & $\underset{\left(195\right)}{9.45}$ & $\underset{\left(358\right)}{9.52}$ & $\underset{\left(678\right)}{9.53}$ & $\underset{\left(1310\right)}{9.53}$ & $\pm0.02$ & $9.55$\tabularnewline
	\bottomrule
	\end{tabular}}
\par\end{centering}
\caption{\label{tab:2}Geo-Call spread, with $\rho_{i,j}=0$ and with $M=2^{d}$. The column labeled \quotes{BM} shows the benchmark obtained via  one-dimensional reduction. \myrb{The column labeled \quotes{ME} gives the margin of error (95\% CI half-width) for NNU results}. Values in brackets indicate computational times (in seconds).}
\end{table}

\medskip
\myrb{Table~\ref{tab:3} reports the performance of GTU and NNU in the pricing of a high-dimensional geometric call spread under zero correlation. To mitigate the exponential growth in GTU's computational cost as the dimension \( d \) increases, we adopt a sampling-based strategy: instead of summing over all \( 2^d \) possible branches, we randomly select a subset of \( M \) representative terms. This Monte Carlo-type simplification is theoretically justified, as the approximated price converges to the full-sum value as \( M \to 2^d \). Empirically, the results in Table~\ref{tab:3} show that even a few thousand samples are sufficient to obtain accurate and stable estimates, enabling the use of GTU for dimensions as high as \( d = 40 \) within reasonable computation times. 
Nevertheless, the GTU approach becomes significantly more costly and less effective for very large dimensions, such as \( d = 80 \). In this regime, we observe both a substantial increase in computational burden and a slower convergence to the benchmark, particularly for coarse discretizations. This illustrates a key limitation of tree-based methods, which are inherently affected by the curse of dimensionality.

By contrast, the NNU method exhibits outstanding stability and accuracy across all tested dimensions. It consistently produces prices that are virtually indistinguishable from the benchmark, with negligible margins of error and computational times that grow only linearly with the number of training epochs. Even for \( d = 80 \), NNU delivers precise results with manageable computational effort, highlighting its scalability and robustness.

In conclusion, while GTU remains effective in low- to moderate-dimensional settings, its performance deteriorates as the dimensionality grows. NNU clearly outperforms GTU in high dimensions, offering a more efficient and practical solution for the valuation of derivatives in large-scale market models.}

\begin{flushleft}
\begin{table}
\begin{centering}
\resizebox{16.4cm}{!}{%
	\setlength{\tabcolsep}{4pt}
\begin{tabular}{ccccccccccccccccccc}
	GTU &  &  & \multicolumn{4}{c}{$N=16$} &  & \multicolumn{4}{c}{$N=32$} &  & \multicolumn{4}{c}{$N=64$} &  & \tabularnewline
	\midrule 
	$d$ & $M$ & $P$ & $125$ & $250$ & $500$ & $1000$ &  & $125$ & $250$ & $500$ & $1000$ &  & $125$ & $250$ & $500$ & $1000$ &  & BM\tabularnewline
	\midrule 
	\multirow{4}{*}{$10$} & $\phantom{0}126$ &  & $\underset{\left(54\right)}{9.46}$ & $\underset{\left(129\right)}{9.49}$ & $\underset{\left(384\right)}{9.51}$ & $\underset{\left(1088\right)}{9.53}$ &  & $\underset{\left(119\right)}{9.45}$ & $\underset{\left(309\right)}{9.48}$ & $\underset{\left(894\right)}{9.51}$ & $\underset{\left(3012\right)}{9.53}$ &  & $\underset{\left(275\right)}{9.44}$ & $\underset{\left(735\right)}{9.47}$ & $\underset{\left(2167\right)}{9.51}$ & $\underset{\left(7219\right)}{9.53}$ &  & \multirow{4}{*}{$9.55$}\tabularnewline
	& $\phantom{0}250$ &  & $\underset{\left(73\right)}{9.45}$ & $\underset{\left(193\right)}{9.48}$ & $\underset{\left(641\right)}{9.50}$ & $\underset{\left(1982\right)}{9.53}$ &  & $\underset{\left(168\right)}{9.44}$ & $\underset{\left(509\right)}{9.47}$ & $\underset{\left(1826\right)}{9.49}$ & $\underset{\left(5350\right)}{9.53}$ &  & $\underset{\left(445\right)}{9.43}$ & $\underset{\left(1172\right)}{9.46}$ & $\underset{\left(4550\right)}{9.49}$ & $\underset{\left(13831\right)}{9.52}$ &  & \tabularnewline
	& $\phantom{0}500$ &  & $\underset{\left(109\right)}{9.45}$ & $\underset{\left(345\right)}{9.47}$ & $\underset{\left(1162\right)}{9.49}$ & $\underset{\left(3808\right)}{9.53}$ &  & $\underset{\left(261\right)}{9.43}$ & $\underset{\left(884\right)}{9.46}$ & $\underset{\left(2942\right)}{9.48}$ & $\underset{\left(10034\right)}{9.53}$ &  & $\underset{\left(785\right)}{9.43}$ & $\underset{\left(2951\right)}{9.45}$ & $\underset{\left(7556\right)}{9.48}$ & $\underset{\left(25161\right)}{9.52}$ &  & \tabularnewline
	& $1000$ &  & $\underset{\left(197\right)}{9.45}$ & $\underset{\left(589\right)}{9.50}$ & $\underset{\left(2006\right)}{9.51}$ & $\underset{\left(6868\right)}{9.54}$ &  & $\underset{\left(465\right)}{9.45}$ & $\underset{\left(1520\right)}{9.49}$ & $\underset{\left(5461\right)}{9.51}$ & $\underset{\left(18118\right)}{9.54}$ &  & $\underset{\left(1360\right)}{9.44}$ & $\underset{\left(3856\right)}{9.48}$ & $\underset{\left(15310\right)}{9.51}$ & $\underset{\left(50299\right)}{9.52}$ &  & \tabularnewline
	\cmidrule{2-19}
	\multirow{4}{*}{$20$} & $\phantom{0}126$ &  & $\underset{\left(57\right)}{9.43}$ & $\underset{\left(140\right)}{9.50}$ & $\underset{\left(420\right)}{9.54}$ & $\underset{\left(1409\right)}{9.57}$ &  & $\underset{\left(145\right)}{9.42}$ & $\underset{\left(359\right)}{9.50}$ & $\underset{\left(1084\right)}{9.54}$ & $\underset{\left(3851\right)}{9.56}$ &  & $\underset{\left(324\right)}{9.42}$ & $\underset{\left(924\right)}{9.50}$ & $\underset{\left(2812\right)}{9.54}$ & $\underset{\left(9659\right)}{9.57}$ &  & \multirow{4}{*}{$9.53$}\tabularnewline
	& $\phantom{0}250$ &  & $\underset{\left(79\right)}{9.43}$ & $\underset{\left(211\right)}{9.50}$ & $\underset{\left(685\right)}{9.55}$ & $\underset{\left(2424\right)}{9.56}$ &  & $\underset{\left(178\right)}{9.43}$ & $\underset{\left(529\right)}{9.50}$ & $\underset{\left(1864\right)}{9.55}$ & $\underset{\left(6560\right)}{9.57}$ &  & $\underset{\left(515\right)}{9.42}$ & $\underset{\left(1426\right)}{9.50}$ & $\text{ \ensuremath{\underset{\left(5481\right)}{9.55}}}$ & $\underset{\left(17040\right)}{9.57}$ &  & \tabularnewline
	& $\phantom{0}500$ &  & $\underset{\left(113\right)}{9.44}$ & $\underset{\left(361\right)}{9.50}$ & $\underset{\left(1185\right)}{9.54}$ & $\underset{\left(4462\right)}{9.56}$ &  & $\underset{\left(262\right)}{9.43}$ & $\underset{\left(956\right)}{9.50}$ & $\underset{\left(3139\right)}{9.54}$ & $\underset{\left(11728\right)}{9.56}$ &  & $\underset{\left(812\right)}{9.43}$ & $\underset{\left(2563\right)}{9.51}$ & $\underset{\left(8471\right)}{9.54}$ & $\underset{\left(31837\right)}{9.56}$ &  & \tabularnewline
	& $1000$ &  & $\underset{\left(175\right)}{9.44}$ & $\underset{\left(619\right)}{9.51}$ & $\underset{\left(2422\right)}{9.56}$ & $\underset{\left(8330\right)}{9.57}$ &  & $\underset{\left(513\right)}{9.43}$ & $\underset{\left(1662\right)}{9.51}$ & $\underset{\left(6004\right)}{9.55}$ & $\underset{\left(22278\right)}{9.56}$ &  & $\underset{\left(1513\right)}{9.43}$ & $\underset{\left(4525\right)}{9.51}$ & $\underset{\left(16695\right)}{9.55}$ & $\underset{\left(60102\right)}{9.55}$ &  & \tabularnewline
	\cmidrule{2-19}
	\multirow{4}{*}{$40$} & $\phantom{0}126$ &  & $\underset{\left(107\right)}{9.33}$ & $\underset{\left(250\right)}{9.40}$ & $\underset{\left(748\right)}{9.48}$ & $\underset{\left(2512\right)}{9.54}$ &  & $\underset{\left(239\right)}{9.31}$ & $\underset{\left(693\right)}{9.39}$ & $\underset{\left(2153\right)}{9.49}$ & $\underset{\left(7021\right)}{9.54}$ &  & $\underset{\left(478\right)}{9.30}$ & $\underset{\left(2051\right)}{9.38}$ & $\underset{\left(5853\right)}{9.49}$ & $\underset{\left(17887\right)}{9.55}$ &  & \multirow{4}{*}{$9.51$}\tabularnewline
	& $\phantom{0}250$ &  & $\underset{\left(145\right)}{9.33}$ & $\underset{\left(400\right)}{9.40}$ & $\underset{\left(1413\right)}{9.48}$ & $\underset{\left(4587\right)}{9.53}$ &  & $\underset{\left(366\right)}{9.31}$ & $\underset{\left(1092\right)}{9.39}$ & $\underset{\left(3634\right)}{9.49}$ & $\underset{\left(12721\right)}{9.54}$ &  & $\underset{\left(701\right)}{9.30}$ & $\underset{\left(3283\right)}{9.38}$ & $\underset{\left(10306\right)}{9.49}$ & $\underset{\left(32959\right)}{9.54}$ &  & \tabularnewline
	& $\phantom{0}500$ &  & $\underset{\left(215\right)}{9.33}$ & $\underset{\left(756\right)}{9.40}$ & $\underset{\left(2361\right)}{9.48}$ & $\underset{\left(9599\right)}{9.53}$ &  & $\underset{\left(616\right)}{9.31}$ & $\underset{\left(2306\right)}{9.39}$ & $\underset{\left(6647\right)}{9.49}$ & $\underset{\left(24170\right)}{9.54}$ &  & $\underset{\left(1536\right)}{9.30}$ & $\underset{\left(5886\right)}{9.38}$ & $\underset{\left(17368\right)}{9.49}$ & $\underset{\left(65290\right)}{9.54}$ &  & \tabularnewline
	& $1000$ &  & $\underset{\left(389\right)}{9.33}$ & $\underset{\left(1392\right)}{9.40}$ & $\underset{\left(4396\right)}{9.48}$ & $\underset{\left(16209\right)}{9.53}$ &  & $\underset{\left(1058\right)}{9.32}$ & $\underset{\left(3997\right)}{9.39}$ & $\underset{\left(13458\right)}{9.49}$ & $\underset{\left(46350\right)}{9.54}$ &  & $\underset{\left(3484\right)}{9.30}$ & $\underset{\left(11411\right)}{9.39}$ & $\underset{\left(36140\right)}{9.49}$ & $\underset{\left(129875\right)}{9.53}$ &  & \tabularnewline
	\cmidrule{2-19}
	\multirow{2}{*}{$80$} & $\phantom{0}126$ &  & $\underset{\left(667\right)}{8.49}$ & $\underset{\left(656\right)}{9.19}$ & $\underset{\left(1944\right)}{9.28}$ & $\underset{\left(6218\right)}{9.39}$ &  & $\underset{\left(1848\right)}{8.34}$ & $\underset{\left(1855\right)}{9.06}$ & $\underset{\left(5671\right)}{9.16}$ & $\underset{\left(21314\right)}{9.27}$ &  & $\underset{\left(1591\right)}{8.23}$ & $\underset{\left(2978\right)}{8.81}$ & $\underset{\left(8573\right)}{8.89}$ & $\underset{\left(53581\right)}{8.96}$ &  & \multirow{2}{*}{$9.51$}\tabularnewline
	& $\phantom{0}250$ &  & $\underset{\left(504\right)}{8.49}$ & $\underset{\left(1057\right)}{9.19}$ & $\underset{\left(3280\right)}{9.28}$ & $\underset{\left(11000\right)}{9.39}$ &  & $\underset{\left(1337\right)}{8.34}$ & $\underset{\left(2875\right)}{9.07}$ & $\underset{\left(10155\right)}{9.17}$ & $\underset{\left(33597\right)}{9.27}$ &  & $\underset{\left(1416\right)}{8.23}$ & $\underset{\left(8366\right)}{8.81}$ & $\underset{\left(26919\right)}{8.89}$ & $\underset{\left(172826\right)}{8.96}$ &  & \tabularnewline
	\midrule
	&  &  &  &  &  &  &  &  &  &  &  &  &  &  &  &  &  & \tabularnewline
	NNU &  &  & \multicolumn{4}{c}{$N=16$} &  & \multicolumn{4}{c}{$N=32$} &  & \multicolumn{4}{c}{$N=64$} &  & \tabularnewline
	\midrule 
	$d$ &  & $E$ & $100$ & $200$ & $400$ & $800$ &  & $100$ & $200$ & $400$ & $800$ &  & $100$ & $200$ & $400$ & $800$ & ME & BM\tabularnewline
	\midrule
	$10$ &  &  & $\underset{\left(47\right)}{9.49}$ & $\underset{\left(82\right)}{9.54}$ & $\underset{\left(152\right)}{9.55}$ & $\underset{\left(309\right)}{9.55}$ &  & $\underset{\left(92\right)}{9.46}$ & $\underset{\left(171\right)}{9.54}$ & $\underset{\left(329\right)}{9.54}$ & $\underset{\left(645\right)}{9.54}$ &  & $\underset{\left(195\right)}{9.45}$ & $\underset{\left(358\right)}{9.52}$ & $\underset{\left(678\right)}{9.53}$ & $\underset{\left(1310\right)}{9.53}$ & $\pm0.02$ & $9.55$\tabularnewline
	$20$ &  &  & $\underset{\left(80\right)}{9.46}$ & $\underset{\left(141\right)}{9.51}$ & $\underset{\left(263\right)}{9.51}$ & $\underset{\left(501\right)}{9.51}$ &  & $\underset{\left(141\right)}{9.49}$ & $\underset{\left(255\right)}{9.52}$ & $\underset{\left(483\right)}{9.53}$ & $\underset{\left(953\right)}{9.53}$ &  & $\underset{\left(258\right)}{9.46}$ & $\underset{\left(463\right)}{9.50}$ & $\underset{\left(876\right)}{9.51}$ & $\underset{\left(1706\right)}{9.51}$ & $\pm0.01$ & $9.53$\tabularnewline
	$40$ &  &  & $\underset{\left(96\right)}{9.49}$ & $\underset{\left(172\right)}{9.51}$ & $\underset{\left(325\right)}{9.51}$ & $\underset{\left(628\right)}{9.51}$ &  & $\underset{\left(172\right)}{9.50}$ & $\underset{\left(316\right)}{9.51}$ & $\underset{\left(603\right)}{9.51}$ & $\underset{\left(1182\right)}{9.51}$ &  & $\underset{\left(346\right)}{9.49}$ & $\underset{\left(638\right)}{9.51}$ & $\underset{\left(1224\right)}{9.51}$ & $\underset{\left(2393\right)}{9.51}$ & $\pm0.01$ & $9.51$\tabularnewline
	$80$ &  &  & $\underset{\left(161\right)}{9.50}$ & $\underset{\left(301\right)}{9.51}$ & $\underset{\left(579\right)}{9.51}$ & $\underset{\left(1138\right)}{9.51}$ &  & $\underset{\left(311\right)}{9.50}$ & $\underset{\left(590\right)}{9.50}$ & $\underset{\left(1144\right)}{9.50}$ & $\underset{\left(2213\right)}{9.50}$ &  & $\underset{\left(523\right)}{9.51}$ & $\underset{\left(1010\right)}{9.51}$ & $\underset{\left(1910\right)}{9.51}$ & $\underset{\left(3696\right)}{9.51}$ & $\pm0.01$ & $9.51$\tabularnewline
	\bottomrule
	\end{tabular}}
\par\end{centering}
\caption{\label{tab:3}Geo-Call spread in high dimension, for $\rho_{i,j}=0$
and $M<2^{d}$ variable. The column labeled \quotes{BM} shows the benchmark obtained via  one-dimensional reduction. \myrb{The column labeled \quotes{ME} gives the margin of error (95\% CI half-width) for NNU results.} Values in brackets indicate computational times (in seconds).}
\end{table}
\par\end{flushleft}

\myrb{Tables~\ref{tab:2} and \ref{tab:3} presented numerical results in the case of uncorrelated underlying assets. To assess the robustness of the GTU and NNU algorithms under correlated market conditions, we conducted a new set of simulations in which all pairwise correlations were fixed at either \( \rho = 0.50 \) or \( \rho = 0.75 \). The corresponding results are reported in Table \ref{tab:4}.

Although no analytical benchmark is available in this setting, both methods display a high degree of numerical stability across all tested configurations. GTU remains accurate and well-behaved up to dimension \( d = 10 \), even under stronger correlation, and shows consistent convergence as the number of paths increases. Similarly, NNU maintains its precision and efficiency, producing prices that evolve smoothly with the number of training epochs and remain robust as both the correlation and the dimension increase.

The good agreement between the two methods, despite the absence of a reference solution, reinforces confidence in the reliability of the computed prices. These findings confirm that both GTU and NNU generalise well to moderately correlated systems, and suggest that their performance remains stable even in the presence of substantial dependence among risk factors.}

\begin{table}
\begin{centering}
\resizebox{16.4cm}{!}{ %
	\setlength{\tabcolsep}{4pt} 
\begin{tabular}{ccccccccccccccccc}
	GTU &  & \multicolumn{4}{c}{$N=16$} &  & \multicolumn{4}{c}{$N=32$} &  & \multicolumn{4}{c}{$N=64$} & \tabularnewline
	\midrule 
	$d$ & $P$ & $125$ & $250$ & $500$ & $1000$ &  & $125$ & $250$ & $500$ & $1000$ &  & $125$ & $250$ & $500$ & $1000$ & \tabularnewline
	\midrule 
	\multicolumn{16}{l}{CASE $\rho_{i,j}=0.50$} & \tabularnewline
	$2$ &  & $\underset{\left(26\right)}{10.90}$ & $\underset{\left(43\right)}{10.89}$ & $\underset{\left(95\right)}{10.88}$ & $\underset{\left(211\right)}{10.88}$ &  & $\underset{\left(46\right)}{10.92}$ & $\underset{\left(91\right)}{10.91}$ & $\underset{\left(181\right)}{10.91}$ & $\underset{\left(459\right)}{10.90}$ &  & $\underset{\left(90\right)}{10.93}$ & $\underset{\left(178\right)}{10.92}$ & $\underset{\left(376\right)}{10.91}$ & $\underset{\left(910\right)}{10.91}$ & \tabularnewline
	$5$ &  & $\underset{\left(30\right)}{10.71}$ & $\underset{\left(57\right)}{10.68}$ & $\underset{\left(131\right)}{10.68}$ & $\underset{\left(343\right)}{10.66}$ &  & $\underset{\left(52\right)}{10.75}$ & $\underset{\left(115\right)}{10.73}$ & $\underset{\left(259\right)}{10.72}$ & $\underset{\left(762\right)}{10.70}$ &  & $\underset{\left(107\right)}{10.75}$ & $\underset{\left(240\right)}{10.75}$ & $\underset{\left(588\right)}{10.74}$ & $\underset{\left(1688\right)}{10.72}$ & \tabularnewline
	$10$ &  & $\underset{\left(150\right)}{10.62}$ & $\underset{\left(532\right)}{10.66}$ & $\underset{\left(1748\right)}{10.62}$ & $\underset{\left(6214\right)}{10.59}$ &  & $\underset{\left(398\right)}{10.64}$ & $\underset{\left(1234\right)}{10.70}$ & $\underset{\left(4284\right)}{10.67}$ & $\underset{\left(16489\right)}{10.64}$ &  & $\underset{\left(906\right)}{10.62}$ & $\underset{\left(2895\right)}{10.72}$ & $\underset{\left(10497\right)}{10.69}$ & $\underset{\left(38313\right)}{10.68}$ & \tabularnewline
	\midrule
	\multicolumn{16}{l}{CASE $\rho_{i,j}=0.75$} & \tabularnewline
	$2$ &  & $\underset{\left(27\right)}{11.02}$ & $\underset{\left(46\right)}{11.05}$ & $\underset{\left(84\right)}{11.05}$ & $\underset{\left(212\right)}{11.04}$ &  & $\underset{\left(44\right)}{11.03}$ & $\underset{\left(82\right)}{11.06}$ & $\underset{\left(175\right)}{11.06}$ & $\underset{\left(441\right)}{11.06}$ &  & $\underset{\left(83\right)}{11.03}$ & $\underset{\left(163\right)}{11.06}$ & $\underset{\left(350\right)}{11.07}$ & $\underset{\left(909\right)}{11.07}$ & \tabularnewline
	$5$ &  & $\underset{\left(28\right)}{11.00}$ & $\underset{\left(51\right)}{10.98}$ & $\underset{\left(115\right)}{10.96}$ & $\underset{\left(347\right)}{10.96}$ &  & $\underset{\left(55\right)}{11.03}$ & $\underset{\left(116\right)}{11.03}$ & $\underset{\left(249\right)}{11.00}$ & $\underset{\left(808\right)}{11.00}$ &  & $\underset{\left(106\right)}{11.04}$ & $\underset{\left(234\right)}{11.06}$ & $\underset{\left(536\right)}{11.02}$ & $\underset{\left(1839\right)}{11.03}$ & \tabularnewline
	$10$ &  & $\underset{\left(129\right)}{11.07}$ & $\underset{\left(491\right)}{11.03}$ & $\underset{\left(1695\right)}{10.98}$ & $\underset{\left(6714\right)}{10.96}$ &  & $\underset{\left(378\right)}{11.10}$ & $\underset{\left(1108\right)}{11.08}$ & $\underset{\left(4126\right)}{11.03}$ & $\underset{\left(15925\right)}{11.02}$ &  & $\underset{\left(816\right)}{11.09}$ & $\underset{\left(2695\right)}{11.10}$ & $\underset{\left(10790\right)}{11.05}$ & $\underset{\left(38559\right)}{11.05}$ & \tabularnewline
	\midrule
	&  &  &  &  &  &  &  &  &  &  &  &  &  &  &  & \tabularnewline
	NNU &  & \multicolumn{4}{c}{$N=16$} &  & \multicolumn{4}{c}{$N=32$} &  & \multicolumn{4}{c}{$N=64$} & \tabularnewline
	\midrule 
	$d$ & $E$ & $100$ & $200$ & $400$ & $800$ &  & $100$ & $200$ & $400$ & $800$ &  & $100$ & $200$ & $400$ & $800$ & ME\tabularnewline
	\midrule
	\multicolumn{16}{l}{CASE $\rho_{i,j}=0.50$} & \tabularnewline
	$2$ &  & $\underset{\left(53\right)}{10.74}$ & $\underset{\left(98\right)}{10.79}$ & $\underset{\left(189\right)}{10.80}$ & $\underset{\left(369\right)}{10.81}$ &  & $\underset{\left(104\right)}{10.79}$ & $\underset{\left(194\right)}{10.88}$ & $\underset{\left(373\right)}{10.89}$ & $\underset{\left(732\right)}{10.90}$ &  & $\underset{\left(196\right)}{10.83}$ & $\underset{\left(366\right)}{10.89}$ & $\underset{\left(706\right)}{10.90}$ & $\underset{\left(1382\right)}{10.91}$ & $\pm0.05$\tabularnewline
	$5$ &  & $\underset{\left(57\right)}{10.62}$ & $\underset{\left(105\right)}{10.65}$ & $\underset{\left(202\right)}{10.66}$ & $\underset{\left(394\right)}{10.66}$ &  & $\underset{\left(130\right)}{10.61}$ & $\underset{\left(246\right)}{10.64}$ & $\underset{\left(475\right)}{10.64}$ & $\underset{\left(961\right)}{10.65}$ &  & $\underset{\left(222\right)}{10.59}$ & $\underset{\left(414\right)}{10.64}$ & $\underset{\left(799\right)}{10.65}$ & $\underset{\left(1568\right)}{10.66}$ & $\pm0.03$\tabularnewline
	$10$ &  & $\underset{\left(61\right)}{10.48}$ & $\underset{\left(121\right)}{10.50}$ & $\underset{\left(240\right)}{10.51}$ & $\underset{\left(477\right)}{10.51}$ &  & $\underset{\left(121\right)}{10.49}$ & $\underset{\left(224\right)}{10.52}$ & $\underset{\left(430\right)}{10.53}$ & $\underset{\left(843\right)}{10.53}$ &  & $\underset{\left(288\right)}{10.53}$ & $\underset{\left(545\right)}{10.57}$ & $\underset{\left(1027\right)}{10.58}$ & $\underset{\left(1903\right)}{10.58}$ & $\pm0.02$\tabularnewline
	\midrule
	\multicolumn{16}{l}{CASE $\rho_{i,j}=0.75$} & \tabularnewline
	$2$ &  & $\underset{\left(54\right)}{10.89}$ & $\underset{\left(99\right)}{10.96}$ & $\underset{\left(190\right)}{10.97}$ & $\underset{\left(371\right)}{10.97}$ &  & $\underset{\left(104\right)}{10.95}$ & $\underset{\left(195\right)}{11.01}$ & $\underset{\left(374\right)}{11.03}$ & $\underset{\left(733\right)}{11.03}$ &  & $\underset{\left(262\right)}{10.94}$ & $\underset{\left(498\right)}{11.03}$ & $\underset{\left(971\right)}{11.05}$ & $\underset{\left(1902\right)}{11.05}$ & $\pm0.05$\tabularnewline
	$5$ &  & $\underset{\left(59\right)}{10.79}$ & $\underset{\left(107\right)}{10.84}$ & $\underset{\left(203\right)}{10.85}$ & $\underset{\left(394\right)}{10.85}$ &  & $\underset{\left(117\right)}{10.83}$ & $\underset{\left(223\right)}{10.87}$ & $\underset{\left(435\right)}{10.88}$ & $\underset{\left(859\right)}{10.89}$ &  & $\underset{\left(219\right)}{10.90}$ & $\underset{\left(410\right)}{10.94}$ & $\underset{\left(792\right)}{10.95}$ & $\underset{\left(1555\right)}{10.95}$ & $\pm0.03$\tabularnewline
	$10$ &  & $\underset{\left(63\right)}{10.80}$ & $\underset{\left(114\right)}{10.83}$ & $\underset{\left(217\right)}{10.84}$ & $\underset{\left(424\right)}{10.84}$ &  & $\underset{\left(124\right)}{10.86}$ & $\underset{\left(231\right)}{10.89}$ & $\underset{\left(442\right)}{10.90}$ & $\underset{\left(864\right)}{10.91}$ &  & $\underset{\left(247\right)}{10.92}$ & $\underset{\left(464\right)}{10.94}$ & $\underset{\left(897\right)}{10.95}$ & $\underset{\left(1764\right)}{10.96}$ & $\pm0.02$\tabularnewline
	\bottomrule
	\end{tabular}}
\par\end{centering}
\caption{\label{tab:4}Geo-Call spread with correlated underlyings.  \myrb{The column labeled \quotes{ME} gives the margin of error (95\% CI half-width) for NNU results.} Values in brackets indicate computational times (in seconds).}
\end{table}

\FloatBarrier
\subsection{Model with uncertainty on correlation}

We now test the GTU algorithm under the assumption that the correlation
coefficients $\rho_{i,j}$ are also unknown, let them vary between
two bounds $\rho_{i,j}^{\min}=-0.5$ and $\rho_{i,j}^{\max}=0.5$.
Evaluating an option in this scenario is more complex for two reasons.
Firstly, it is necessary to verify that the correlation matrix $\Gamma$,
resulting from varying the coefficients $\rho_{i,j}$ during the optimization
procedure, remains positive semidefinite. \myrb{As for GTU, the verification is performed directly using \textsc{Matlab}'s \texttt{chol} function, whereas for NNU, we follow the penalization approach described in Section~\ref{sec:NNCU}}. Secondly, the number
of these coefficients increases quadratically with dimension, leading
to a rapid increase in computational cost, making it challenging to
obtain results in high dimensions.

\subsubsection{Outperformer spread option with variable correlation}

The payoff of this option is given in (\ref{eq:Outperformer_spread}).
We only consider the case $d=2$ since this is the only case considered
by \cite{guyon2010uncertain}. Moreover, in this particular case it
is unnecessary to verify that the correlation matrix $\Gamma$ is
positive semidefinite, as this holds true for any permissible value
of $\rho_{1,2}$. We propose this particular case because it has also
been studied by \cite{guyon2010uncertain} and thus it is possible
to test our algorithm against theirs. In this regard, we can observe
that our results reported in Table 7 are slightly higher than the
reference value in \cite{guyon2010uncertain}. This slight difference
probably results from the fact that our optimization procedure performs
better than Guyon and Labordère's one in \cite{guyon2010uncertain}
and therefore, it is better able to approach the optimal price in
UVM. We also note that the price obtained with a variable correlation
parameter, specifically $-0.5\leq\rho_{1,2}\leq0.5$, is significantly
higher than the prices obtained considering a fixed correlation of
$\rho_{1,2}=0.5$, $\rho_{1,2}=0$ or $\rho_{1,2}=-0.5$, shown in
Table \ref{tab:6}. Specifically, the higher result for a fixed correlation
is achieved for $\rho_{1,2}=-0.5$ and it is equal to $11.39$, while
for a variable correlation, we reach $12.77$ \myrb{ 	with GTU and $12.80\pm 0.05$ with NNU.} 

\myrb{ 	
	We point out that GTU delivers accurate results even with limited computational effort. In particular, for \( N = 32 \) and \( P = 250 \), GTU achieves a price of $12.72$ -- very close to the reference value of $12.67$ in under 300 seconds. 
	By comparison, NNU reaches similar levels of accuracy only at higher training costs: the same value is approximated with \( N=64 \) and \(E = 400 \) demanding a computation time of about $1000$ seconds. This highlights the competitive performance of GTU in low-dimensional settings, especially when rapid estimation is needed.
	
}

To conclude, it is
crucial to emphasize the importance of considering a robust model
for correlation among the underlying assets, as pricing results can
vary significantly. Therefore, variable correlation is a risk factor
that practitioners should not overlook.
\begin{center}
\begin{table}
\begin{centering}
\scalebox{0.9}{ 
\begin{tabular}{ccccccc}
	\multicolumn{7}{l}{GTU}\tabularnewline
	\midrule
	$N/P$ & $125$ & $250$ & $500$ & $1000$ & $2000$ & \tabularnewline
	\midrule
	$32$ & $\underset{\left(129\right)}{12.60}$ & $\underset{\left(242\right)}{12.72}$ & $\underset{\left(577\right)}{12.72}$ & $\underset{\left(1754\right)}{12.73}$ & $\underset{\left(2302\right)}{12.74}$ & \tabularnewline
	$64$ & $\underset{\left(380\right)}{12.59}$ & $\underset{\left(537\right)}{12.73}$ & $\underset{\left(1244\right)}{12.74}$ & $\underset{\left(3490\right)}{12.76}$ & $\underset{\left(4791\right)}{12.77}$ & \tabularnewline
	$128$ & $\underset{\left(355\right)}{12.52}$ & $\underset{\left(733\right)}{12.69}$ & $\underset{\left(1466\right)}{12.72}$ & $\underset{\left(4116\right)}{12.75}$ & $\underset{\left(10331\right)}{12.77}$ & \tabularnewline
	\midrule
	&  &  &  &  &  & \tabularnewline
	\multicolumn{7}{l}{NNU}\tabularnewline
	\midrule
	$N/E$ & $100$ & $200$ & $400$ & $800$ & $1600$ & ME\tabularnewline
	\midrule
	$32$ & $\underset{\left(144\right)}{11.32}$ & $\underset{\left(259\right)}{12.49}$ & $\underset{\left(491\right)}{12.61}$ & $\underset{\left(962\right)}{12.64}$ & $\underset{\left(1905\right)}{12.64}$ & $\pm0.05$\tabularnewline
	$64$ & $\underset{\left(276\right)}{11.09}$ & $\underset{\left(506\right)}{12.48}$ & $\underset{\left(968\right)}{12.70}$ & $\underset{\left(1890\right)}{12.74}$ & $\underset{\left(3721\right)}{12.76}$ & $\pm0.05$\tabularnewline
	$128$ & $\underset{\left(634\right)}{11.11}$ & $\underset{\left(1176\right)}{12.58}$ & $\underset{\left(2301\right)}{12.76}$ & $\underset{\left(4591\right)}{12.79}$ & $\underset{\left(9180\right)}{12.80}$ & $\pm0.05$\tabularnewline
	\bottomrule
	\end{tabular}}
\par\end{centering}
\caption{Outperformer spread option with variable correlation $\rho_{1,2}$.
The value computed by   \cite{guyon2010uncertain} for the same option is $12.67$.  \myrb{The column labeled \quotes{ME} gives the margin of error (95\% CI half-width) for NNU results.} Values in brackets indicate computational times (in seconds).}
\end{table}
\par\end{center}

\subsubsection{Geo-Outperformer option with variable correlation}

In Subsection \ref{subsec:Outperformer-option} we have introduced
the Outperformer option, in which the payoff (\ref{eq:outperformer})
depends only on 2 underlyings. In this Subsection, we extend the definition
of such an option to an option on multiple underlyings. Specifically,
we replace the second underlying with the geometric mean of all the
underlying with the exception of the first one, so the payoff function
reads out:

\begin{equation}
\Psi\left(\mathbf{S}_{T}\right)=\left(\sqrt[d-1]{\prod_{k=2}^{d}S_{T}^{k}}-S_{T}^{1}\right)_{+}.\label{eq:geo_out}
\end{equation}

In this particular case, the optimal combination of the $\rho_{i,j}$
coefficients is easily found: maximize the correlation between all
the underlyings except the first one and minimize the correlation
between the first underlying and the remaining ones. In our numerical
setting, this implies
\begin{equation}
\rho_{i,j}=\begin{cases}
	1 & \text{if } i = j, \\
	-0.5 & \text{if } i \neq j \text{ and } \min(i,j) = 1, \\
	0.5 & \text{otherwise}
\end{cases}\label{eq:coef}
\end{equation}
This solutions, in fact, maximizes the volatility of the difference
in (\ref{eq:geo_out}) and thus the expected value of the payoff.
This remark allows us to compute a benchmark by means of the GTU algorithm
for correlation fixed according to (\ref{eq:coef}). In particular,
we compute the Benchmark by using $N=128$ time steps and $P=1000$
points.

The results presented in Table \ref{tab:8} are generally consistent
with the benchmark, albeit slightly lower \myrb{for both the two algorithms}. \myrb{In this regard, the less accurate final result in the table is obtained by NNU in the $5$-dimensional case, where the final value of $12.45$ falls below the benchmark of $12.64$, corresponding to a relative error of \(-1.5\%\). Nonetheless, this still represents an entirely acceptable level of accuracy.
} \myrb{As far as GTU is considered,} this discrepancy is likely
due to the method used to construct the sets $X_{n}$ required for
GPR. Specifically, as far as the benchmark is considered, these sets
are constructed using optimal correlation in (\ref{eq:coef}), whereas
for the general problem with variable correlation, they are determined
by using the average correlation (\ref{eq:rho_avg}), which is sub-optimal.
Consequently, more points are needed, compared to the fixed correlation
model, to achieve results of equivalent quality.

\myrb{As for NNU, the observed underperformance is likely due to the computational complexity of the problem. For instance, in the case \( d = 5 \), the network is required to output $5$ volatilities and $10$ correlations, for a total of $15$ heterogeneous parameters at each time step. Increasing the network's capacity may help improve accuracy in this setting; however, this comes at the cost of longer computational times and it does not seem to be convenient given the quality of the proposed results.
}
\begin{center}
\begin{table}
\begin{centering}
\resizebox{16.4cm}{!}{\setlength{\tabcolsep}{3pt}
\begin{tabular}{ccccccccccccccccccccc}
	GTU &  & \multicolumn{5}{c}{$N=32$} &  & \multicolumn{5}{c}{$N=64$} &  & \multicolumn{5}{c}{$N=128$} &  & BM\tabularnewline
	\midrule 
	$d$ & $P$ & $125$ & $250$ & $500$ & $1000$ & $2000$ &  & $125$ & $250$ & $500$ & $1000$ & $2000$ &  & $125$ & $250$ & $500$ & $1000$ & $2000$ &  & \tabularnewline
	\midrule
	$2$ &  & $\underset{\left(70\right)}{13.78}$ & $\underset{\left(138\right)}{13.79}$ & $\underset{\left(304\right)}{13.78}$ & $\underset{\left(941\right)}{13.78}$ & $\underset{\left(2491\right)}{13.78}$ &  & $\underset{\left(146\right)}{13.76}$ & $\underset{\left(301\right)}{13.78}$ & $\underset{\left(719\right)}{13.77}$ & $\underset{\left(2127\right)}{13.76}$ & $\underset{\left(6283\right)}{13.76}$ &  & $\underset{\left(312\right)}{13.74}$ & $\underset{\left(671\right)}{13.78}$ & $\underset{\left(1548\right)}{13.77}$ & $\underset{\left(4840\right)}{13.76}$ & $\underset{\left(12463\right)}{13.76}$ &  & $13.75$\tabularnewline
	$3$ &  & $\underset{\left(168\right)}{12.84}$ & $\underset{\left(368\right)}{13.00}$ & $\underset{\left(758\right)}{12.98}$ & $\underset{\left(1982\right)}{12.96}$ & $\underset{\left(4335\right)}{12.95}$ &  & $\underset{\left(421\right)}{12.82}$ & $\underset{\left(768\right)}{12.99}$ & $\underset{\left(1554\right)}{12.96}$ & $\underset{\left(4350\right)}{12.94}$ & $\underset{\left(9336\right)}{12.93}$ &  & $\underset{\left(900\right)}{12.78}$ & $\underset{\left(1633\right)}{12.97}$ & $\underset{\left(3154\right)}{12.96}$ & $\text{ \ensuremath{\underset{\left(8221\right)}{12.94}}}$ & $\underset{\left(17646\right)}{12.93}$ &  & $12.96$\tabularnewline
	$4$ &  & $\underset{\left(384\right)}{12.46}$ & $\underset{\left(716\right)}{12.64}$ & $\underset{\left(1165\right)}{12.67}$ & $\underset{\left(2632\right)}{12.71}$ & $\underset{\left(7216\right)}{12.70}$ &  & $\underset{\left(820\right)}{12.37}$ & $\underset{\left(1349\right)}{12.60}$ & $\underset{\left(2086\right)}{12.64}$ & $\underset{\left(5156\right)}{12.70}$ & $\underset{\left(11520\right)}{12.69}$ &  & $\underset{\left(1833\right)}{12.25}$ & $\underset{\left(2630\right)}{12.53}$ & $\underset{\left(4083\right)}{12.60}$ & $\underset{\left(10019\right)}{12.69}$ & $\underset{\left(22713\right)}{12.69}$ &  & $12.73$\tabularnewline
	$5$ &  & $\underset{\left(2006\right)}{12.15}$ & $\underset{\left(3431\right)}{12.52}$ & $\underset{\left(2022\right)}{12.53}$ & $\underset{\left(6328\right)}{12.62}$ & $\underset{\left(9796\right)}{12.57}$ &  & $\underset{\left(1847\right)}{12.05}$ & $\underset{\left(2646\right)}{12.45}$ & $\underset{\left(3918\right)}{12.49}$ & $\underset{\left(10362\right)}{12.60}$ & $\underset{\left(18192\right)}{12.56}$ &  & $\underset{\left(3534\right)}{11.90}$ & $\underset{\left(5103\right)}{12.35}$ & $\underset{\left(7420\right)}{12.41}$ & $\underset{\left(20298\right)}{12.57}$ & $\underset{\left(54465\right)}{12.54}$ &  & $12.64$\tabularnewline
	\midrule
	&  &  &  &  &  &  &  &  &  &  &  &  &  &  &  &  &  &  &  & \tabularnewline
	NNU &  & \multicolumn{5}{c}{$N=32$} &  & \multicolumn{5}{c}{$N=64$} &  & \multicolumn{5}{c}{$N=128$} & ME & BM\tabularnewline
	\midrule 
	$d$ & $E$ & $100$ & $200$ & $400$ & $800$ & $1600$ &  & $100$ & $200$ & $400$ & $800$ & $1600$ &  & $100$ & $200$ & $400$ & $800$ & $1600$ &  & \tabularnewline
	\midrule
	$2$ &  & $\underset{\left(163\right)}{13.33}$ & $\underset{\left(302\right)}{13.70}$ & $\underset{\left(581\right)}{13.74}$ & $\underset{\left(1139\right)}{13.75}$ & $\underset{\left(2255\right)}{13.75}$ &  & $\underset{\left(279\right)}{13.19}$ & $\underset{\left(513\right)}{13.73}$ & $\underset{\left(981\right)}{13.78}$ & $\underset{\left(1916\right)}{13.78}$ & $\underset{\left(3788\right)}{13.79}$ &  & $\underset{\left(592\right)}{13.10}$ & $\underset{\left(1087\right)}{13.65}$ & $\underset{\left(2077\right)}{13.69}$ & $\underset{\left(4061\right)}{13.70}$ & $\underset{\left(7832\right)}{13.70}$ & $\pm0.13$ & $13.75$\tabularnewline
	$3$ &  & $\underset{\left(271\right)}{12.32}$ & $\underset{\left(511\right)}{12.78}$ & $\underset{\left(992\right)}{12.83}$ & $\underset{\left(1947\right)}{12.83}$ & $\underset{\left(3805\right)}{12.84}$ &  & $\underset{\left(626\right)}{12.16}$ & $\underset{\left(1189\right)}{12.77}$ & $\underset{\left(2306\right)}{12.83}$ & $\underset{\left(4535\right)}{12.84}$ & $\underset{\left(8982\right)}{12.84}$ &  & $\underset{\left(1064\right)}{12.16}$ & $\underset{\left(2016\right)}{12.77}$ & $\underset{\left(3901\right)}{12.81}$ & $\underset{\left(7649\right)}{12.83}$ & $\underset{\left(15116\right)}{12.84}$ & $\pm0.12$ & $12.96$\tabularnewline
	$4$ &  & $\underset{\left(327\right)}{11.57}$ & $\underset{\left(620\right)}{12.59}$ & $\underset{\left(1197\right)}{12.67}$ & $\underset{\left(2344\right)}{12.68}$ & $\underset{\left(4636\right)}{12.68}$ &  & $\underset{\left(647\right)}{11.87}$ & $\underset{\left(1230\right)}{12.50}$ & $\underset{\left(2381\right)}{12.55}$ & $\underset{\left(4669\right)}{12.56}$ & $\underset{\left(9234\right)}{12.56}$ &  & $\underset{\left(1280\right)}{11.68}$ & $\underset{\left(2444\right)}{12.56}$ & $\underset{\left(4743\right)}{12.61}$ & $\underset{\left(9350\right)}{12.62}$ & $\underset{\left(18552\right)}{12.62}$ & $\pm0.11$ & $12.73$\tabularnewline
	$5$ &  & $\underset{\left(469\right)}{11.59}$ & $\underset{\left(898\right)}{12.33}$ & $\underset{\left(1746\right)}{12.37}$ & $\underset{\left(3435\right)}{12.38}$ & $\underset{\left(6798\right)}{12.38}$ &  & $\underset{\left(1057\right)}{11.68}$ & $\underset{\left(2040\right)}{12.40}$ & $\underset{\left(3980\right)}{12.45}$ & $\underset{\left(7826\right)}{12.46}$ & $\underset{\left(15486\right)}{12.46}$ &  & $\underset{\left(1839\right)}{11.83}$ & $\underset{\left(3539\right)}{12.40}$ & $\underset{\left(6920\right)}{12.44}$ & $\underset{\left(13656\right)}{12.45}$ & $\underset{\left(26901\right)}{12.45}$ & $\pm0.11$ & $12.64$\tabularnewline
	\bottomrule
	\end{tabular}}
\par\end{centering}
\caption{\label{tab:8}Outperformer geo option with variable correlation. Values
in brackets represent the computational times, measured in seconds.
The column labeled \quotes{BM} shows
the benchmark computed by using the GTU algorithm with the optimal
correlation parameters.   \myrb{The column labeled \quotes{ME} gives the margin of error (95\% CI half-width) for NNU results.} Values in brackets indicate computational times (in seconds).}
\end{table}
\par\end{center}

\subsection{Path dependent option: the Call Sharpe option}

We conclude our numerical analysis by discussing an option known as
Call Sharpe, previously analyzed by \cite{guyon2010uncertain}. This
option is particularly interesting as it exemplifies a path-dependent
option, where the value depends not only on the current price of the
underlying asset but also on the previously realized volatility. Specifically,
the payoff at maturity, which depends on a single underlying asset,
and on its realized volatility which is computed on a monthly base.
To this aim, we assume that the maturity time $T$ is a multiple of
$1/12$, defining $N_{m}=12T\in\mathbb{N}$ as the number of months
comprising the life of the contract. The realized volatility computed
using monthly returns is given by
\[
V_{T}=\frac{1}{T}\sum_{l=1}^{N_{m}}\left(\ln\left(\frac{S_{\tau_{l}}}{S_{\tau_{l-1}}}\right)\right)^{2},
\]
with $\tau_{l}=l/12$, and the payoff at maturity is
\[
\Psi\left(S_{T}\right)=\frac{\left(S_{T}-K\right)^{+}}{\sqrt{V_{T}}}.
\]
Following \cite{guyon2010uncertain}, at time $t$, the option value
depends on $\ensuremath{S_{t}}$ and on the two path-dependent variables

\begin{align*}
A_{t}^{1} & =\sum_{\{l\mid\tau_{l}\leq t\}}\left(\ln\left(\frac{S_{\tau_{l}}}{S_{\tau_{l-1}}}\right)\right)^{2},\\
A_{t}^{2} & =S_{\sup_{\{l\mid\tau_{l}\leq t\}}\tau_{l}}.
\end{align*}

\myrb{At the monthly monitoring dates \( t = \tau_l \), it holds by construction that \( A^2_t = S_t \). As a result, the variable \( A^2_t \) provides no additional information beyond what is already captured by \( S_t \), and can thus be considered redundant. Therefore, the pricing problem at such times can be formulated as a function of only two variables -- \( S_t \) and \( A^1_t \) -- instead of three. This dimensionality reduction simplifies the regression task and improves computational efficiency.
} 

Since this option is path dependent, some adaptations are required
to use GTU. Let us begin by pointing out that the number of time steps
$N$ must be a multiple of $N_{m}$, so that the monitoring dates
$\tau_{l}$ are included in the time grid.

Next, we note that the elements of the sets $X_{n}$, used to define
the points at which to evaluate the contract value, are determined
through Monte Carlo simulations. Specifically, first we simulate $P$
randoms paths of the underlying stock price:
\begin{equation}
\left\{ S_{t_{n}}^{p}=S_{t_{n-1}}^{p}e^{\left(r-\eta-\left(\sigma^{\text{avg}}\right)^{2}/2\right)\Delta t+\sigma^{\text{avg}}\sqrt{\Delta t}Z^{n,p}},n=1,\dots,N,p=1,\dots,P\right\} ,\label{eq:Sharpe_P}
\end{equation}
with $S_{0}^{p}=S_{0}$, $\Delta t=T/N$ and $Z^{n,p}\sim\mathcal{N}\left(0,1\right)$.
Then, the elements $\mathbf{x}^{n,p}$ of $X_{n}$ are defined as
\[
\mathbf{x}^{n,p}=\begin{cases}
\left(S_{t_{n}}^{p},A_{t_{n}}^{1,p}\right) & \text{if }12t_{n}\in\mathbb{N},\\
\left(S_{t_{n}}^{p},A_{t_{n}}^{1,p},A_{t_{n}}^{2,p}\right) & \text{otherwise},
\end{cases}
\]
with
\[
A_{t_{n}}^{1,p}=\sum_{\{l\mid\tau_{l}\leq t_{n}\}}\left(\ln\left(\frac{S_{\tau_{l}}^{p}}{S_{\tau_{l-1}}^{p}}\right)\right)^{2},\quad A_{t_{n}}^{2,p}=S_{\sup_{\{l\mid\tau_{l}\leq t\}}\tau_{l}}^{p}.
\]
We emphasize  that, as far as this path dependent option is considered,
we opt for the Monte Carlo method over the quasi-Monte Carlo method
to generate the points $\mathbf{x}^{n,p}$ of $X_{n}$ in (\ref{eq:Sharpe_P}),
as the dimension of random simulations equals $N_{m}$, to much to
make quasi-Monte Carlo be effective with only a few hundreds of samples.

For a particular value $\hat{\sigma}$ for the volatility, determined
during the optimization procedure, the future points for $\mathbf{x}^{n,p}$
are only 2, namely $\tilde{\mathbf{x}}_{i}^{n,p,1}$ and $\tilde{\mathbf{x}}_{i}^{n,p,2}$,
defined as
\[
\tilde{\mathbf{x}}_{i}^{n,p,m}=\begin{cases}
\left(S_{t_{n}}^{p,m},A_{t_{n}}^{1,p}+\left(\ln\left(\frac{S_{t_{n}}^{p,m}}{A_{t_{n}}^{2}}\right)\right)^{2}\right) & \text{if }12t_{n+1}\in\mathbb{N},\\
\left(S_{t_{n}}^{p,m},A_{t_{n}}^{1,p}+\left(\ln\left(\frac{S_{t_{n}}^{p,m}}{A_{t_{n}}^{2}}\right)\right)^{2},A_{t_{n}}^{2,p}\right) & \text{otherwise}.
\end{cases}
\]
for $m=1,2$ and with 
\[
S_{t_{n}}^{p,m}=S_{t_{n}}^{p}e^{\left(r-\eta-\left(\hat{\sigma}\right)^{2}/2\right)\Delta t+\hat{\sigma}\sqrt{\Delta t}\left(-1\right)^{m}}.
\]
Moreover, as far as this path dependent option is considered, the
elements that identify the market state, namely $S,A^{1}$ and $A^{2}$,
represent very different quantities from each other. For this reason,
as remarked in a similar framework by \cite{goudenege2020machine},
it is beneficial to use a kernel that considers a different length
scale for each predictor, such as the ARD Matérn 3/2 kernel $k_{\text{M3/2}}^{\text{ARD}}$
which is defines as follows:
\[
k_{\text{M3/2}}^{\text{ARD}}(\mathbf{x},\mathbf{x}')=\sigma_{f}^{2}\left(1+\sqrt{3\sum_{i}\left(\frac{\mathbf{x}_{i}-\mathbf{x}'_{i}}{\sigma_{l,i}}\right)^{2}}\right)\exp\left(-\frac{\sqrt{3}}{\sigma_{l}}\sqrt{3\sum_{i}\left(\frac{\mathbf{x}_{i}-\mathbf{x}'_{i}}{\sigma_{l,i}}\right)^{2}}\right),
\]
where $\sigma_{l,i}$ is the length scale for the $i$-th predictor.

\myrb{Adaptations similar to those discussed for GTU are also necessary for NNU. In addition, given the computational complexity of the problem under consideration, we increased the number of neurons for each layer from $32$ to $64$.
Moreover, although the number of independent inputs is reduced to three at monitoring times, we choose to always feed the network with $5$ inputs, namely   time  $t_n$, intra-interval position $\sup_{\{l\mid\tau_{l}\leq t\}}\tau_{l}$ (i.e., the time elapsed since the last monitoring date), current price $S_n$, accumulated variance $A^1_{t_n}$, and the last monitored price $A^2_{t_n}$. This design ensures that the network architecture remains uniform across all time steps. 
}

In the specific case study considered in our numerical test, we consider
$T=1$, so $N_{m}=12$. Results are reported in Table (\ref{tab:9}). 
The benchmark value, computed through a PDE approach described in
\cite{guyon2010uncertain}, is $58.40$. \\
\myrb{For GTU, the most accurate result is obtained with \( N = 384 \) and \( P = 2000 \), yielding a price of 57.93, which is much closer to the benchmark than the best value reported in \cite{guyon2010uncertain} (55.55). The relative deviation is below 1\%, indicating a high level of accuracy. This test case also confirms that a large number of time steps \( N \) is needed to attain accurate results, thus highlighting the importance of using a low-cost algorithm per time step. The GTU method, which in this case evaluates only two future values per point, proves particularly efficient in this regard.
	
	As for NNU, the results show consistent improvement as both the number of time steps \( N \) and the number of training epochs \( E \) increase. Although the final estimates remain slightly below the PDE benchmark, the accuracy improves steadily with training, reaching $57.08\pm 0.44$ for \( N = 384 \) and \( E = 3200 \), with a relative deviation of approximately 2.3\%. While NNU requires longer computational times than GTU to reach this level of precision, it remains competitive and exhibits stable convergence.
	
	In conclusion, GTU achieves more accurate results at lower computational cost for this problem. However, NNU remains a valid alternative, particularly when flexibility and model generalisation are prioritized, and its performance continues to improve with deeper training.}
 
 \begin{rem} 
 	\myrb{The convergence of both  {GTU} and  {NNU} is primarily influenced by two factors: the number of time steps $N$ used in the discretisation of the control problem, and the size of the statistical model (either the number of training points $P$ for kriging in  {GTU}, or the number of epochs $E$ in the training of the neural net in  {NNU}).
 		
 In the case of  {GTU}, increasing $N$ improves the resolution of the dynamic programming recursion and reduces the discretisation bias. For path-independent options, numerical tests show that results become stable already at $N \approx 64$ (for instance, in the Geo-Call spread, the estimated price varies by less than $0.04$ between $N = 16$ and $N = 64$; see Table \ref{tab:2}). For path-dependent payoffs, such as the Call-Sharpe contract, a finer time grid is needed, typically $N \geq 256$, to obtain results that match the PDE benchmark within a fraction of a percent (see Table \ref{tab:9}). Similar behaviour is observed with  {NNU}, where path-independent contracts require a coarse grid only, while path-dependent options benefit from a finer temporal resolution.
 		
 The second key component is the statistical approximation. In  {GTU}, the Gaussian process surrogate is based on a Matérn $3/2$ kernel, whose mean-square error decreases with the number of training points $P$, depending on the input dimension and the smoothness of the target. Empirically, we observe that $P = 500$ to $1000$ suffices to make the regression error negligible compared to the discretisation bias. In the NNU, the accuracy depends instead on the number of training epochs $E$. Due to the expressive power of feedforward neural nets \cite{hornik1991approximation} and their effectiveness in hedging applications \cite{Buehler2019}, the pricing error decreases rapidly with $E$. In the Geo-Call example, the NNU estimate lies within the Monte Carlo confidence band after $E \approx 400$, and stabilises around the benchmark value by $E = 800$.
 		 
 	}
 \end{rem}
 
\begin{center}
\begin{table}
\begin{centering}
\scalebox{0.9}{\setlength{\tabcolsep}{6pt}%

\begin{tabular}{cccccccccccc}
	& \multicolumn{4}{c}{GTU} &  &  & \multicolumn{4}{c}{NNU} & \tabularnewline
	\cmidrule{1-5}\cmidrule{7-12}
	$N/P$ & $250$ & $500$ & $1000$ & $2000$ & \hspace{9mm} & $N/E$ & $400$ & $800$ & $1600$ & $3200$ & ME\tabularnewline
	\cmidrule{1-5}\cmidrule{7-12}
	$24$ & $\underset{\left(59\right)}{55.97}$ & $\underset{\left(109\right)}{55.94}$ & $\underset{\left(373\right)}{56.04}$ & $\underset{\left(1325\right)}{56.35}$ &  & $24$ & $\underset{\left(299\right)}{51.77}$ & $\underset{\left(590\right)}{52.64}$ & $\underset{\left(1171\right)}{52.91}$ & $\underset{\left(2333\right)}{52.97}$ & $\pm0.42$\tabularnewline
	$48$ & $\underset{\left(98\right)}{59.35}$ & $\underset{\left(217\right)}{57.26}$ & $\underset{\left(751\right)}{57.14}$ & $\underset{\left(2030\right)}{57.32}$ &  & $48$ & $\underset{\left(707\right)}{53.36}$ & $\underset{\left(1405\right)}{54.32}$ & $\underset{\left(2802\right)}{54.58}$ & $\underset{\left(5592\right)}{54.62}$ & $\pm0.42$\tabularnewline
	$96$ & $\underset{\left(199\right)}{57.74}$ & $\underset{\left(556\right)}{57.71}$ & $\underset{\left(1580\right)}{57.46}$ & $\underset{\left(4494\right)}{57.66}$ &  & $96$ & $\underset{\left(1277\right)}{53.71}$ & $\underset{\left(1277\right)}{53.71}$ & $\underset{\left(5060\right)}{55.54}$ & $\underset{\left(10103\right)}{55.60}$ & $\pm0.42$\tabularnewline
	$192$ & $\underset{\left(386\right)}{59.83}$ & $\underset{\left(1016\right)}{57.30}$ & $\underset{\left(3233\right)}{58.19}$ & $\underset{\left(10114\right)}{57.90}$ &  & $192$ & $\underset{\left(2821\right)}{55.14}$ & $\underset{\left(5628\right)}{55.98}$ & $\underset{\left(11240\right)}{56.27}$ & $\underset{\left(21109\right)}{56.32}$ & $\pm0.43$\tabularnewline
	$384$ & $\underset{\left(872\right)}{58.19}$ & $\underset{\left(2566\right)}{58.93}$ & $\underset{\left(7315\right)}{58.22}$ & $\underset{\left(26228\right)}{57.93}$ &  & $384$ & $\underset{\left(6722\right)}{55.18}$ & $\underset{\left(12768\right)}{56.41}$ & $\text{\ensuremath{\underset{\left(12520\right)}{56.66}}}$ & $\underset{\left(24569\right)}{57.08}$ & $\pm0.44$\tabularnewline
	\cmidrule{1-5}\cmidrule{7-12}
	\end{tabular}}
\par\end{centering}
\caption{\label{tab:9}Call Sharpe. The value computed by \cite{guyon2010uncertain} for the same option is $55.55$, while the benchmark obtained using a PDE approach is $58.40$.  \myrb{The column labeled \quotes{ME} gives the margin of error (95\% CI half-width) for NNU results.} Values in brackets indicate computational times (in seconds).}
\end{table}
\par\end{center}

\FloatBarrier
\section{Conclusions}
\myrb{In this study, we demonstrated the efficacy of Machine Learning in enhancing the Uncertain Volatility Model   for high-dimensional option pricing. Our findings indicate  that both algorithms proposed -- GTU  and NNU -- provide accurate and reliable results across a wide range of settings.

The NNU method proves particularly effective in high-dimensional problems, where its scalability and data-driven flexibility allow it to deliver accurate prices along with confidence intervals, which are valuable for risk assessment and regulatory reporting. Its ability to learn complex, non-linear relationships makes it especially suitable when the dimensionality   the model increases.

In contrast, GTU remains more efficient in low-dimensional settings and in scenarios with uncertain correlation. Its tree-based structure ensures fast and reliable estimates with limited computational cost, making it an ideal choice when rapid evaluation is needed and the model dimensionality is moderate.

Overall, both methods contribute meaningfully to the development of data-driven superhedging strategies within the Uncertain Volatility Model framework. The integration of Machine Learning not only improves the precision of volatility and correlation control, but also enhances computational efficiency, providing robust tools for traders and risk managers. Future research may extend this work by investigating alternative learning architectures or hybrid approaches, further bridging the gap between theoretical models and practical implementations in quantitative finance.
}

\subsubsection*{Acknowledgement}
This work was supported by the Departmental Strategic Plan (PSD) of the University of Udine. Interdepartmental Project on Artificial Intelligence (2020-2025).

\subsubsection*{Declaration of generative AI and AI-assisted technologies in the writing process}
During the preparation of this work the authors used ChatGPT (version 4o) in order to improve the grammatical quality and
readability. After using this service, the authors reviewed and edited the content as needed and take full responsibility
for the content of the publication.

\bibliographystyle{apalike}
\bibliography{UVM_biblio}

\begin{thebibliography}{}

\bibitem[Akhtari et~al., 2023]{akhtari2023generalized}
Akhtari, B., Biagini, F., Mazzon, A., and Oberpriller, K. (2023).
\newblock Generalized {Feynman}--{Kac} formula under volatility uncertainty.
\newblock {\em Stochastic Processes and their Applications}, 166:104083.

\bibitem[Andersen and Broadie, 2004]{andersen2004primal}
Andersen, L. and Broadie, M. (2004).
\newblock Primal-dual simulation algorithm for pricing multidimensional
  {American} options.
\newblock {\em Management Science}, 50(9):1222--1234.

\bibitem[Avellaneda et~al., 1995]{avellaneda1995pricing}
Avellaneda, M., Levy, A., and Paras, A. (1995).
\newblock Pricing and hedging derivative securities in markets with uncertain
  volatilities.
\newblock {\em Applied Mathematical Finance}, 2(2):73--88.

\bibitem[Becker et~al., 2019]{becker2019deep}
Becker, S., Cheridito, P., and Jentzen, A. (2019).
\newblock Deep optimal stopping.
\newblock {\em Journal of Machine Learning Research}, 20:1--25.

\bibitem[Becker et~al., 2020]{becker2020pricing}
Becker, S., Cheridito, P., and Jentzen, A. (2020).
\newblock Pricing and hedging {{American}}-style options with {Deep Learning}.
\newblock {\em Journal of Risk and Financial Management}, 13(7):158.

\bibitem[Becker et~al., 2021]{becker2021solving}
Becker, S., Cheridito, P., Jentzen, A., and Welti, T. (2021).
\newblock Solving high-dimensional optimal stopping problems using {Deep
  Learning}.
\newblock {\em European Journal of Applied Mathematics}, 32(3):470--514.

\bibitem[Biggs, 1975]{biggs1975constrained}
Biggs, M. (1975).
\newblock Constrained minimization using recursive quadratic programming.
\newblock {\em Towards global optimization}.

\bibitem[Buehler et~al., 2019]{Buehler2019}
Buehler, H., Gonon, L., Teichmann, J., and Wood, B. (2019).
\newblock Deep hedging.
\newblock {\em Quantitative Finance}, 19(8):1271--1291.

\bibitem[Calafiore and El~Ghaoui, 2014]{calafiore2014optimization}
Calafiore, G. and El~Ghaoui, L. (2014).
\newblock {\em Optimization Models}.
\newblock Cambridge University Press, Cambridge.

\bibitem[Cohen and Tegn{\'e}r, 2019]{cohen2019european}
Cohen, S.~N. and Tegn{\'e}r, M. (2019).
\newblock European option pricing with stochastic volatility models under
  parameter uncertainty.
\newblock In {\em Frontiers in Stochastic Analysis--BSDEs, SPDEs and their
  Applications: Edinburgh, July 2017 Selected, Revised and Extended
  Contributions 8}, pages 123--167. Springer.

\bibitem[De~Spiegeleer et~al., 2018]{de2018machine}
De~Spiegeleer, J., Madan, D.~B., Reyners, S., and Schoutens, W. (2018).
\newblock Machine learning for quantitative finance: fast derivative pricing,
  hedging and fitting.
\newblock {\em Quantitative Finance}, 18(10):1635--1643.

\bibitem[E et~al., 2017]{EHanJentzen2017}
E, W., Han, J., and Jentzen, A. (2017).
\newblock Deep learning-based numerical methods for high-dimensional parabolic
  partial differential equations and backward stochastic differential
  equations.
\newblock {\em Communications in Mathematics and Statistics}, 5(4):349--380.

\bibitem[Ekvall, 1996]{ekvall1996lattice}
Ekvall, N. (1996).
\newblock A lattice approach for pricing of multivariate contingent claims.
\newblock {\em European Journal of Operational Research}, 91(2):214--228.

\bibitem[Fletcher, 2013]{fletcher2013practical}
Fletcher, R. (2013).
\newblock {\em Practical Methods of Optimization}.
\newblock John Wiley \& Sons, Hoboken, NJ.

\bibitem[Gouden{\`e}ge et~al., 2020]{goudenege2020machine}
Gouden{\`e}ge, L., Molent, A., and Zanette, A. (2020).
\newblock Machine learning for pricing american options in high-dimensional
  markovian and non-markovian models.
\newblock {\em Quantitative Finance}, 20(4):573--591.

\bibitem[Gouden{\`e}ge et~al., 2021]{goudenege2021gaussian}
Gouden{\`e}ge, L., Molent, A., and Zanette, A. (2021).
\newblock Gaussian process regression for pricing variable annuities with
  stochastic volatility and interest rate.
\newblock {\em Decisions in Economics and Finance}, 44(1):57--72.

\bibitem[Guyon and Henry-Labord{\`e}re, 2011]{guyon2010uncertain}
Guyon, J. and Henry-Labord{\`e}re, P. (2011).
\newblock Uncertain volatility model: a {Monte Carlo} approach.
\newblock {\em Journal of Computational Finance}, 14(3):37--71.

\bibitem[Han et~al., 2018]{EHanJentzen2018}
Han, J., Jentzen, A., and Weinan, E. (2018).
\newblock Solving high-dimensional partial differential equations using deep
  learning.
\newblock {\em Proceedings of the National Academy of Sciences},
  115(34):8505--8510.

\bibitem[Han, 1977]{han1977globally}
Han, S.-P. (1977).
\newblock A globally convergent method for nonlinear programming.
\newblock {\em Journal of optimization theory and applications},
  22(3):297--309.

\bibitem[Haugh and Kogan, 2004]{haugh2004pricing}
Haugh, M.~B. and Kogan, L. (2004).
\newblock Pricing {American} options: A duality approach.
\newblock {\em Operations Research}, 52(2):258--270.

\bibitem[Hornik, 1991]{hornik1991approximation}
Hornik, K. (1991).
\newblock Approximation capabilities of multilayer feedforward networks.
\newblock {\em Neural networks}, 4(2):251--257.

\bibitem[Kamath et~al., 2018]{kamath2018neural}
Kamath, A., Vargas-Hern{\'a}ndez, R.~A., Krems, R.~V., Carrington, T., and
  Manzhos, S. (2018).
\newblock Neural networks vs {Gaussian} process regression for representing
  potential energy surfaces: A comparative study of fit quality and vibrational
  spectrum accuracy.
\newblock {\em The Journal of chemical physics}, 148(24).

\bibitem[Lapeyre and Lelong, 2021]{lapeyre2021neural}
Lapeyre, B. and Lelong, J. (2021).
\newblock Neural network regression for {Bermudan} option pricing.
\newblock {\em Monte Carlo Methods and Applications}, 27(3):227--247.

\bibitem[Longstaff and Schwartz, 2001]{longstaff2001valuing}
Longstaff, F.~A. and Schwartz, E.~S. (2001).
\newblock Valuing {American} options by simulation: a simple least-squares
  approach.
\newblock {\em The Review of Financial Studies}, 14(1):113--147.

\bibitem[Ludkovski, 2018a]{ludkovski2018}
Ludkovski, M. (2018a).
\newblock {Kriging} metamodels and experimental design for {Bermudan} option
  pricing.
\newblock {\em Journal of Computational Finance}, 22(1):37--77.

\bibitem[Ludkovski, 2018b]{ludkovski2018kriging}
Ludkovski, M. (2018b).
\newblock Kriging metamodels and experimental design for bermudan option
  pricing.
\newblock {\em Journal of Computational Finance}, 22(1).

\bibitem[L{\"u}tkebohmert et~al., 2022]{lutkebohmert2022robust}
L{\"u}tkebohmert, E., Schmidt, T., and Sester, J. (2022).
\newblock Robust deep hedging.
\newblock {\em Quantitative Finance}, 22(8):1465--1480.

\bibitem[Mallick et~al., 2021]{mallick2021deep}
Mallick, A., Dwivedi, C., Kailkhura, B., Joshi, G., and Han, T. Y.-J. (2021).
\newblock Deep kernels with probabilistic embeddings for small-data learning.
\newblock In {\em Uncertainty in artificial intelligence}, pages 918--928.
  PMLR.

\bibitem[Margrabe, 1978]{margrabe1978value}
Margrabe, W. (1978).
\newblock The value of an option to exchange one asset for another.
\newblock {\em The journal of finance}, 33(1):177--186.

\bibitem[Martini and Jacquier, 2008]{martini2008uncertain}
Martini, C. and Jacquier, A. (2008).
\newblock The uncertain volatility model.
\newblock Working Paper.

\bibitem[{MathWorks}, 2024]{mathworksKernelCovariance}
{MathWorks} (2024).
\newblock Kernel covariance function options - {MATLAB Statistics and Machine
  Learning Toolbox}.
\newblock
  \url{https://it.mathworks.com/help/stats/kernel-covariance-function-options.html}.
\newblock Accessed: 2025-05-22.

\bibitem[MathWorks{\,}Inc., 2024]{MathWorks2024}
MathWorks{\,}Inc., T. (2024).
\newblock {\em {\normalfont \textsc{MATLAB}} Statistics and Machine Learning
  Toolbox}.
\newblock Function \texttt{fitrgp}: Gaussian Process Regression.

\bibitem[Nadarajah et~al., 2017]{nadarajah2017comparison}
Nadarajah, S., Margot, F., and Secomandi, N. (2017).
\newblock Comparison of least squares {Monte Carlo} methods with applications
  to energy real options.
\newblock {\em European Journal of Operational Research}, 256(1):196--204.

\bibitem[Nocedal and Wright, 2006]{nocedal2006numerical}
Nocedal, J. and Wright, S. (2006).
\newblock {\em Numerical Optimization}.
\newblock Springer Series in Operations Research and Financial Engineering.
  Springer New York.

\bibitem[Pooley et~al., 2003]{pooley2003numerical}
Pooley, D.~M., Gerrard, C., and Muir, D.~M. (2003).
\newblock Numerical analysis of the uncertain volatility model.
\newblock {\em Applied Mathematical Finance}, 10(3):211--240.

\bibitem[Powell, 1978]{Powell1978}
Powell, M.~J. (1978).
\newblock The convergence of variable metric methods for nonlinearly
  constrained optimization calculations.
\newblock In {\em Nonlinear programming 3}, pages 27--63. Elsevier, New York,
  San Francisco, London.

\bibitem[Powell, 2006]{Powell2006}
Powell, M.~J. (2006).
\newblock A fast algorithm for nonlinearly constrained optimization
  calculations.
\newblock In {\em Numerical Analysis: Proceedings of the Biennial Conference
  Held at Dundee, June 28--July 1, 1977}, pages 144--157. Springer.

\bibitem[Ruppert and Matteson, 2011]{ruppert2011statistics}
Ruppert, D. and Matteson, D.~S. (2011).
\newblock {\em Statistics and Data Analysis for Financial Engineering},
  volume~13.
\newblock Springer, New York.

\bibitem[Schittkowski, 1986]{Schittkowski1986}
Schittkowski, K. (1986).
\newblock {NLPQL}: A {FORTRAN} subroutine solving constrained nonlinear
  programming problems.
\newblock {\em Annals of operations research}, 5:485--500.

\bibitem[Williams and Rasmussen, 2006]{williams2006gaussian}
Williams, C. K.~I. and Rasmussen, C.~E. (2006).
\newblock {\em Gaussian Processes for Machine Learning}.
\newblock MIT Press, Cambridge, MA.

\bibitem[Windcliff et~al., 2006]{windcliff2006numerical}
Windcliff, H., Forsyth, P., and Vetzal, K. (2006).
\newblock Numerical methods and volatility models for valuing cliquet options.
\newblock {\em Applied Mathematical Finance}, 13(4):353--386.

\end{thebibliography}

\end{document}